\documentclass[12pt,superscriptaddress,nature]{revtex4-2}
\usepackage{soul}
\usepackage{orcidlink}
\usepackage{graphicx}
\usepackage{epstopdf}
\usepackage[utf8]{inputenc}
\usepackage{amsmath,amsthm,amssymb}
\usepackage{latexsym}
\usepackage{bm}
\usepackage{dcolumn}
\usepackage{braket}
\usepackage[normalem]{ulem}
\usepackage{times}
\usepackage{hyperref}
\usepackage[section]{placeins}

\newcommand{\ba}{\mbox{\boldmath$a$}}

\newcommand{\bJ}{\mbox{\boldmath$J$}}

\newcommand{\s}{\sigma}

  \def\func#1{{\rm \expandafter\string#1}}
  \def\re{\ensuremath{{\func{Re}}\,}}
  
  \def\iu{\ensuremath{\mathrm{i}\mkern1mu}}

  \def\ERF{\ensuremath{{\func{ERF}}\,}}

  \newcommand{\Eq}[1]{Eq.~\eqref{eq:#1}}

  \newcommand{\Fig}[1]{Fig.~\ref{#1}}

\begin{document}

\title{High-temperature quantum coherence of spinons in an Yb spin chain}
\title{High-temperature quantum coherence of an Yb spin chain}
\title{High-temperature quantum coherence of a rare-earth spin chain}
\title{Observation of hot fractionalized quantum matter}
\title{Hot quantum matter in a rare-earth spin chain}
\title{High-temperature quantum spinons in a rare-earth perovskite crystal}
\title{Hot quantum spinons in a rare-earth spin chain}
\title{Hot quantum spinons in a rare-earth perovskite crystal}
\title{Hot Fermi gas of quantum spinons in a rare-earth perovskite crystal}
\title{High-temperature quantum coherence of spinons in a rare-earth perovskite}
\title{High-temperature quantum coherence of spinons in a rare-earth spin chain}

\author{Lazar L. Kish \orcidlink{0000-0001-7132-8415} }
\affiliation{Condensed Matter Physics and Materials Science Division, Brookhaven National Laboratory, Upton, NY 11973, USA}

\author{Andreas Weichselbaum \orcidlink{0000-0002-5832-3908} }
\affiliation{Condensed Matter Physics and Materials Science Division, Brookhaven National Laboratory, Upton, NY 11973, USA}

\author{Daniel M. Pajerowski \orcidlink{0000-0003-3890-2379} }
\affiliation{Neutron Scattering Division, Oak Ridge National Laboratory, Oak Ridge, TN 37831, USA}

\author{Andrei T. Savici \orcidlink{0000-0001-5127-8967} }
\affiliation{Neutron Scattering Division, Oak Ridge National Laboratory, Oak Ridge, TN 37831, USA}

\author{Andrey Podlesnyak \orcidlink{0000-0001-9366-6319} }
\affiliation{Neutron Scattering Division, Oak Ridge National Laboratory, Oak Ridge, TN 37831, USA}

\author{Leonid Vasylechko \orcidlink{0000-0003-4231-9186} }
\affiliation{Lviv Polytechnic National University, Lviv, Ukraine}

\author{Alexei Tsvelik \orcidlink{0000-0002-7478-670X} }
\affiliation{Condensed Matter Physics and Materials Science Division, Brookhaven National Laboratory, Upton, NY 11973, USA}

\author{Robert Konik \orcidlink{0000-0003-1209-6890} }
\affiliation{Condensed Matter Physics and Materials Science Division, Brookhaven National Laboratory, Upton, NY 11973, USA}

\author{Igor A. Zaliznyak \orcidlink{0000-0002-8548-7924} }
\email{zaliznyak@bnl.gov}
\affiliation{Condensed Matter Physics and Materials Science Division, Brookhaven National Laboratory, Upton, NY 11973, USA}

\begin{abstract}
{\bf
Conventional wisdom dictates that quantum effects become unimportant at high temperatures. In magnets, when the thermal energy exceeds interactions between atomic magnetic moments, the moments are usually uncorrelated, and classical paramagnetic behavior is observed. This thermal decoherence of quantum spin behaviors is a major hindrance to quantum information applications of spin systems. Remarkably, our neutron scattering experiments on Yb chains in an insulating perovskite crystal defy these conventional expectations. We find a sharply defined spectrum of spinons, fractional quantum excitations of spin-1/2 chains, to persist to temperatures much higher than the scale of the interactions between Yb magnetic moments. The observed sharpness of the spinon continuum's dispersive upper boundary indicates a spinon mean free path exceeding $\approx 35$ inter-atomic spacings at temperatures more than an order of magnitude above the interaction energy scale. We thus discover an important and highly unique quantum behavior, which expands the realm of quantumness to high temperatures where entropy-governed classical behaviors were previously believed to dominate. Our results have profound implications for spin systems in quantum information applications operating at finite temperatures and motivate new developments in quantum metrology.
}

\bigskip
\emph{One sentence summary:}
Our neutron scattering study reveals a remarkable persistence of quantum spin coherence in an Yb qubit candidate material at temperatures more than an order of magnitude above the energy scale of interactions governing its spin dynamics.

\end{abstract}

\date{\today}

\maketitle
\newpage

Magnetism is the oldest quantum phenomenon, known for nearly 2500 years before it was understood following the discovery of electron spin \cite{Uhlenbeck_Nature1926} and the invention of quantum mechanics \cite{Dirac_book}. Beyond simple ferromagnetism, quantum theory predicts a great variety of other collective spin states, such as in exactly solved antiferromagnetic spin-1/2 chains \cite{Bethe_1931}, where spins exhibit long-range quantum entanglement but no static magnetic order. Consequently, spin systems are widely considered for quantum information applications requiring quantum-coherent processing, transmission, and storage of entangled states. Quantum computation and communication algorithms using spin chains \cite{Bose_PRL2003,CamposVenuti_PRL2007,Tserkovnyak_PRA2011,Marchukov_NatComm2016,Thompson_NJP2016}, fractional and topological excitations in quantum spin liquids \cite{Kitaev_AnnPhys2003,Broholm_Science2020,Semeghini_Science2021}, as well as magnons in ordered ferro- and antiferro-magnets \cite{Andrich_NPJQM2017,LachanceQuirion_2019,Chumak_2022} are currently being investigated.

The main hurdle for quantum computing applications is a decoherence of entangled states when unwanted interactions with the environment or thermal excitations cause quantum information to be lost. The long-range coherence of quantum states existing at zero temperature, $T = 0$, can be destroyed at $T>0$ when excitations change their identities by colliding and exchanging quantum numbers, as is seen in the thermal decoherence of phonon-roton excitations in superfluid helium \cite{Nichitiu_PRB2024}. In a quantum spin-1 chain, where the Haldane ground state is disordered, magnon excitations are separated from it by an energy gap, $\Delta_H$, and exhibit mesoscopic long-range coherence at $T = 0$ \cite{Xu_Science2007}. However, coherence is rapidly lost as magnons become thermally excited at temperatures $k_BT \sim \Delta_H$ ($k_B$ is Boltzmann constant) \cite{Zaliznyak_PRB1994,Xu_Science2007,Zheludev_PRL2008}. Such decoherence of the magnon excitations which encode quantum states can be described as a finite collisional lifetime, which in this case can be accurately calculated \cite{Sachdev_PRL1997}. A similar phenomenology can be seen in ordered magnets as well, where magnons become over-damped, entirely losing their coherent quantum nature as the thermal energy becomes comparable to magnon bandwidths \cite{Huberman_2008,Bayrakci_PRL2013}. This thermal decoherence limits the potential applications of magnons for the storage and transmission of quantum information.

Here, we find an entirely different situation in the case of spinons, fractional excitations in a spin-1/2 chain. Our magnetic inelastic neutron scattering (INS) measurements show that in a material realization of spin-1/2 chains in the rare earth perovskite YbAlO$_3$ \cite{Wu_PRB2019,Wu_2019,Nikitin_PRB2020} spinons retain their quantum coherence to temperatures where thermal energy exceeds characteristic energy scales of spin interactions by more than an order of magnitude. Moreover, an eventual reduction of quantum coherence at our highest measured temperatures stems from interactions with high-energy thermal bath-type degrees of freedom external to the effective spin Hamiltonian.

\begin{figure*}[t!h!]
\includegraphics[width=0.75\textwidth]{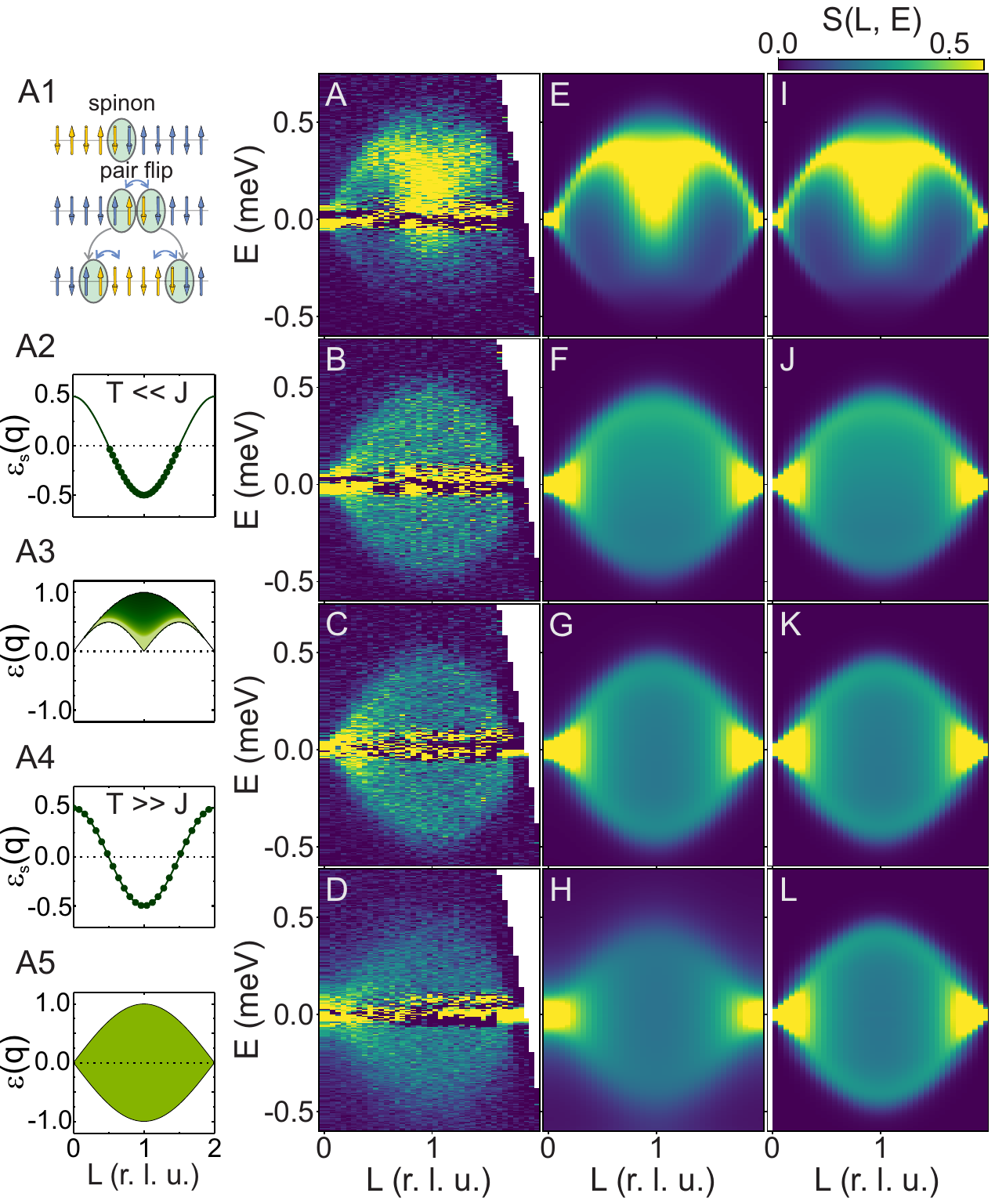}
\caption{{\bf The spinon spectra in YbAlO$_{3}$ at different temperatures. } (A1) Schematic illustration of spinons as topological defects in antiferromagnetic spin chain; spinon pairs measured in our INS experiments are created by pairwise nearest-spin flips. (A2-A5) schematics of how half-filled fermion band gives rise to two-spinon continuum boundaries (see also Supplementary Information \cite{SI}). (A-D) Color contour maps of the spectral density of the measured neutron scattering intensity at different temperatures. These data are integrated in the dispersionless transverse directions with $K=[-1.0, 1.0]$ and $H=[-0.25, 0.25]$. (E-H) Fits to model constructed from DMRG calculations with Lorentzian broadening accounting for spinon lifetime, as reported in the main text, directly comparable to neutron data. (I-L) Resolution-corrected DMRG calculations without additional Lorentzian broadening accounting for spinon finite lifetime for comparison.}
\label{Fig1:Fig1_Exp_Fit_DMRG}
\end{figure*}

The magnetic doublets of rare earth Kramers ions such as Yb$^{3+}$ in a crystal electric field (CEF) provide a fruitful approach to implementing quantum spin qubits in solids \cite{Awschalom_2018,Zhong_2017,Ruskuc_Nature2022,Beckert_NatPhys2024}. Although such a doublet has orbital character imposed by a strong spin-orbit coupling (SOC), it can be represented as a pseudo-spin-1/2, similar to the real spin-1/2 of an unpaired magnetic electron, implementing a quantum qubit. Advantageously, the states of a doublet can carry large angular momentum quantum numbers, which suppresses their interaction with magnetic fields of the environment by virtue of selection rules expressing angular momentum conservation \cite{Dirac_book}. Hence, rare earth spin qubits can have longer coherence times \cite{Zhong_2017,Ruskuc_Nature2022}. Such is the situation of Yb$^{3+}$ ions in YbAlO$_3$ \cite{Wu_PRB2019,Wu_2019,Nikitin_PRB2020,Nikitin_NatCom2021}, which we study here. Strong SOC (one of the strongest among all lanthanides) combines the spin ($S = 1/2$) and the orbital ($L = 3$) angular momenta of a single hole in the 4$f$ shell of Yb$^{3+}$ into a total angular momentum \bJ\ ($J = 7/2$) state, effectively quenching the spin degree of freedom by rigidly tying it to the dominant orbital contribution. This leads to a very simple electronic level structure, which is within the reach of near-infrared or visible photons. Consequently, Yb atoms make the world's most accurate atomic clocks, highly efficient high-power crystal and fiber lasers and optical amplifiers, 
and are a promising system for optically controlled quantum information applications \cite{Zhong_2017}. A chain of coupled Yb spins (doublets) in YbAlO$_3$ implements a chain of coupled spin qubits where coherently propagating spinon excitations act to switch the state of each qubit, as illustrated in Fig.~\ref{Fig1:Fig1_Exp_Fit_DMRG}(A1).

While spin-1/2 chains in magnetic crystals have been studied in the past \cite{Zaliznyak_PRL2004,Zaliznyak_NatMat2005,Lake_NatMat2005,Mourigal_NatPhys2013}, to our knowledge the important question of what happens to spinon excitations at high temperature remains experimentally unexplored. This is largely because the exchange energy scales in most studied spin-chain materials are in the range of tens to hundreds of meV ($J/k_B \sim 100-1000$~K, $k_B$ is Boltzmann constant), which makes it difficult to reach temperatures truly in excess of the interaction energies. From this perspective, YbAlO$_3$ is an ideal material to study because of the relatively weak exchange interaction in its effective spin-1/2 Hamiltonian ($J \approx 0.21$~meV, $J/k_B \approx 2.4$~K \cite{Wu_2019,Nikitin_PRB2020}) and an absence of magnetic order down to a temperature of 0.8~K. As a result, we can use inelastic neutron spectroscopy to probe the physics of the Heisenberg spin-1/2 chain in a temperature regime that is unattainable in other spin-1/2 chain materials.

Here, we report a detailed INS investigation of the spinon spectrum in YbAlO$_3$ as a function of temperature in the $2-100$~K ($\sim (1 - 40) \times J$) range. The excitation spectrum of the ideal spin-1/2 Heisenberg chain is known to consist of pairs of spinons, fractional elementary excitations each carrying S = 1/2 angular momentum \cite{Bethe_1931}. Pair-states of these spinon excitations encode physical spin flips in the chain [this is schematically illustrated in Fig.~\ref{Fig1:Fig1_Exp_Fit_DMRG}(A1)], whose energy spectrum forms a continuum, at zero-temperature sharply bounded by the two-spinon boundaries ($q$ is the wave vector, $d$ is the lattice spacing) \cite{Zaliznyak_PRL2004,Zaliznyak_NatMat2005,Lake_NatMat2005,Mourigal_NatPhys2013,Caux_JStatMech2006},
\begin{equation}
\frac{\pi}{2} J|\sin{q d}| \leq \epsilon(q) \leq \pi J \left| \sin{\left( \frac{q d}{2} \right)} \right|
\end{equation}

Qualitatively, the lower and upper two-spinon continuum boundaries show different behavior as a function of temperature, which can be understood by considering spinons as fermion quasiparticles half-filling the one-dimensional energy band, $\epsilon_s (q) = \frac{\pi}{2} J \sin{q d}$ \cite{Gannon_NatComm2019,Nikitin_NatCom2021,SI}. 
The lower continuum boundary arises because of the complete occupation of states below the spinon Fermi energy at zero temperature, which forbids excitations into the filled states [Fig.~\ref{Fig1:Fig1_Exp_Fit_DMRG}(A2,A3)]. With the increasing temperature, the Fermi distribution smears out, allowing state occupations above the Fermi level at the expense of the occupied states below it [Fig.~\ref{Fig1:Fig1_Exp_Fit_DMRG}(A4)]. As a result, the lower boundary blurs until it completely disappears at temperatures $\gtrsim \frac{\pi}{2} J /k_B$. On the other hand, the upper boundary reflects the maximum energy that a spinon pair with a given $q$ can have according to the dispersion, $\epsilon_s (q)$. In the absence of spinon decoherence through a finite collisional lifetime in the idealized system described by the quantum spin-chain Hamiltonian, the profile of the upper two-spinon boundary must remain completely untouched by temperature effects [Fig.~\ref{Fig1:Fig1_Exp_Fit_DMRG}(A5)]. The upper boundary of the excitation continuum is only blurred beyond the two-spinon boundary by the presence of multi-spinon-excitations. At $T = 0$, the total spectral weight above this upper boundary from such excitations is relatively small ($\sim 1\%$) \cite{Caux_JStatMech2006}. While this blurring is in fact temperature-dependent, it is entirely governed by the quantum spin Hamiltonian and as our theoretical calculations show remains insignificant even at high temperatures, $T \gg J/k_B$.

In the presence of couplings to a system external to the quantum spin Hamiltonian, such as a thermal heat bath or other extrinsic source of decoherence, a quantum spin-chain will experience information loss to these external degrees of freedom. This will be reflected by a reduced spinon lifetime, measurable in neutron spectra by a broadening along the energy direction beyond instrument resolution. A blurring of the upper boundary of the excitation continuum in excess of the theoretically calculated width generated by multi-spinon excitations is then a metric for spinon decoherence, quantifying the degree of information loss from the spin-chain to the environment.

\begin{figure*}[t!h!]
\includegraphics[width=0.8\textwidth]{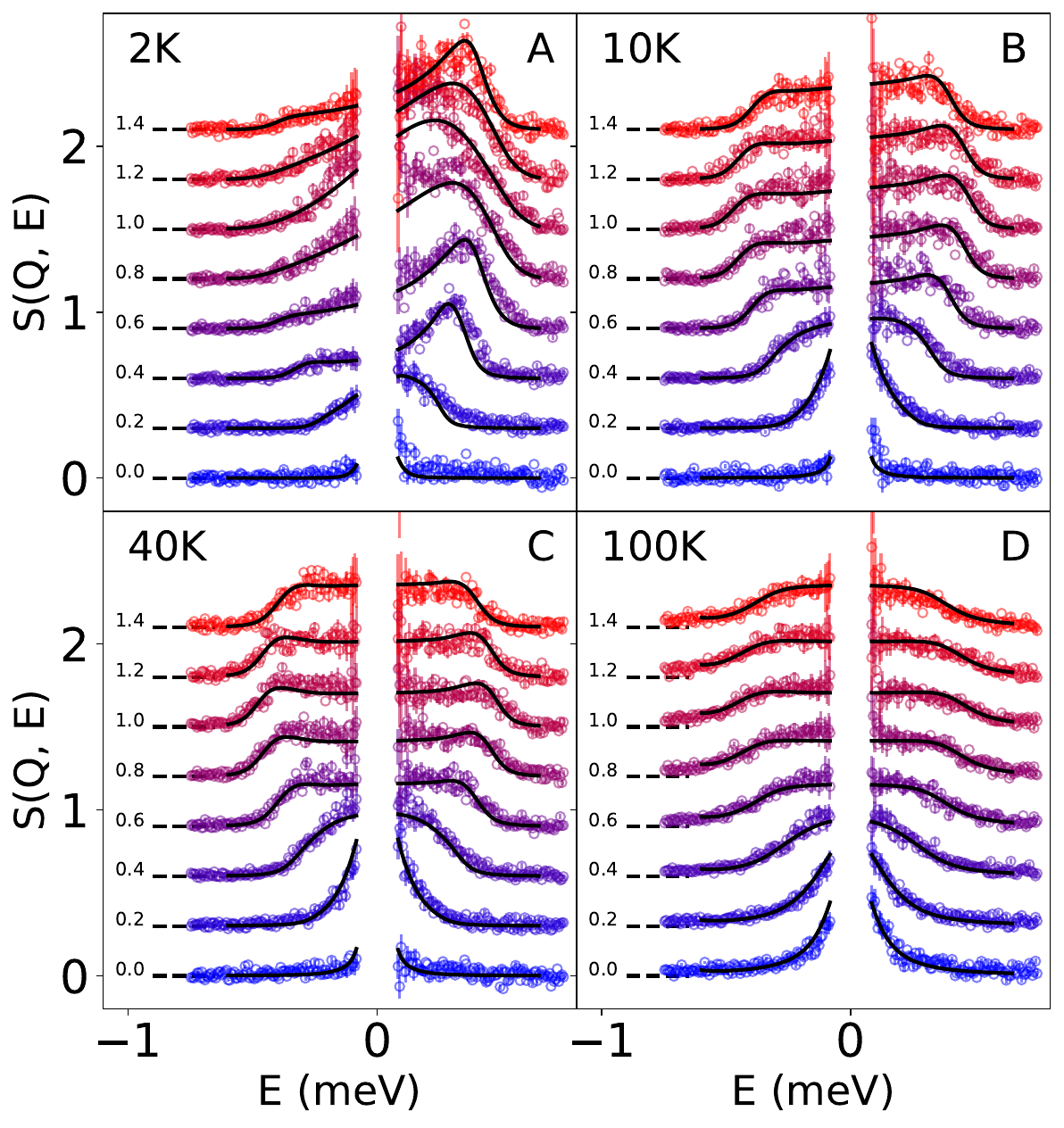}
\caption{{\bf Line cuts along the energy axis of our data and fits to our model. } Curves are given an incremental offset for visualization, with dashed leader-lines from each curve signifying the zero of intensity. The labels next to each curve signify the central L value of each line-cut, which are 0.2 r. l. u. wide. (A)~2~K~(r.~$\chi^2 = 2.1$); (B)~10~K~(r.~$\chi^2 = 1.9$); (C)~40~K~(r.~$\chi^2 = 1.3$); (D)~100~K (r.~$\chi^2 = 1.2$)}
\label{Fig2:waterfall_fit}
\end{figure*}

Figure~\ref{Fig1:Fig1_Exp_Fit_DMRG} shows the temperature dependence of the measured spinon continuum in YbAlO$_3$ side-by-side with temperature-dependent realizations of the spin-1/2 Heisenberg model from finite-temperature DMRG calculations (see Methods). The left column shows our experimentally measured dynamical structure factors, normalized to absolute units as described in the Supplementary Information \cite{SI}. The middle column shows a fit of our DMRG-calculated spectrum to the experimental data, including convolution with the known instrumental resolution function and a Lorentzian broadening function with half-width $\Gamma$ to model finite spinon life-time, $\tau = \hbar / \Gamma$ \cite{Nichitiu_PRB2024,Zaliznyak_PRB1994}. The right column shows the DMRG calculations without the Lorentzian broadening, demonstrating how the spectrum would appear if the effects of spinon thermal decoherence were absent. The waterfall plot in Fig.~\ref{Fig2:waterfall_fit} shows constant-$L$ line-cuts of data and the corresponding Lorentzian-broadened DMRG calculation at selected wave-vectors, which demonstrates the excellent agreement between our model and data (values for the reduced $\chi^2$ goodness-of-fit parameter are listed in the caption and are below 3 for all temperatures; $L$ is the component of the wave vector, $Q = (H, K, L)$, along the chain direction, see Methods).

At 2~K, the lower continuum boundary is visible in both experiment and DMRG simulations, albeit already slightly blurred by thermal repopulation as temperature is comparable to the exchange coupling, $J/k_B = 2.4$~K. At higher temperatures, 10~K and above, all signs of the lower continuum boundary have disappeared in both experiment and simulation and instead been replaced by a flat continuum. This flat continuum, however, remains clearly bounded by the dispersive upper boundary even at temperatures far higher than the exchange coupling. Remarkably, our experimentally measured datasets demonstrate this clear upper-boundary dispersion at temperatures as high as 100~K, forty times greater than the exchange interactions within the system. Only a slight blurring of the upper boundary can be seen, which is most clearly visible in the 1-dimensional plots in Fig.~\ref{Fig2:waterfall_fit}. This blurring appears well modelled by the wave-vector-independent Lorentzian damping, $\Gamma$, indicating finite spinon lifetime at high temperatures.

\begin{figure*}[t!h!]
\includegraphics[width=0.8\textwidth]{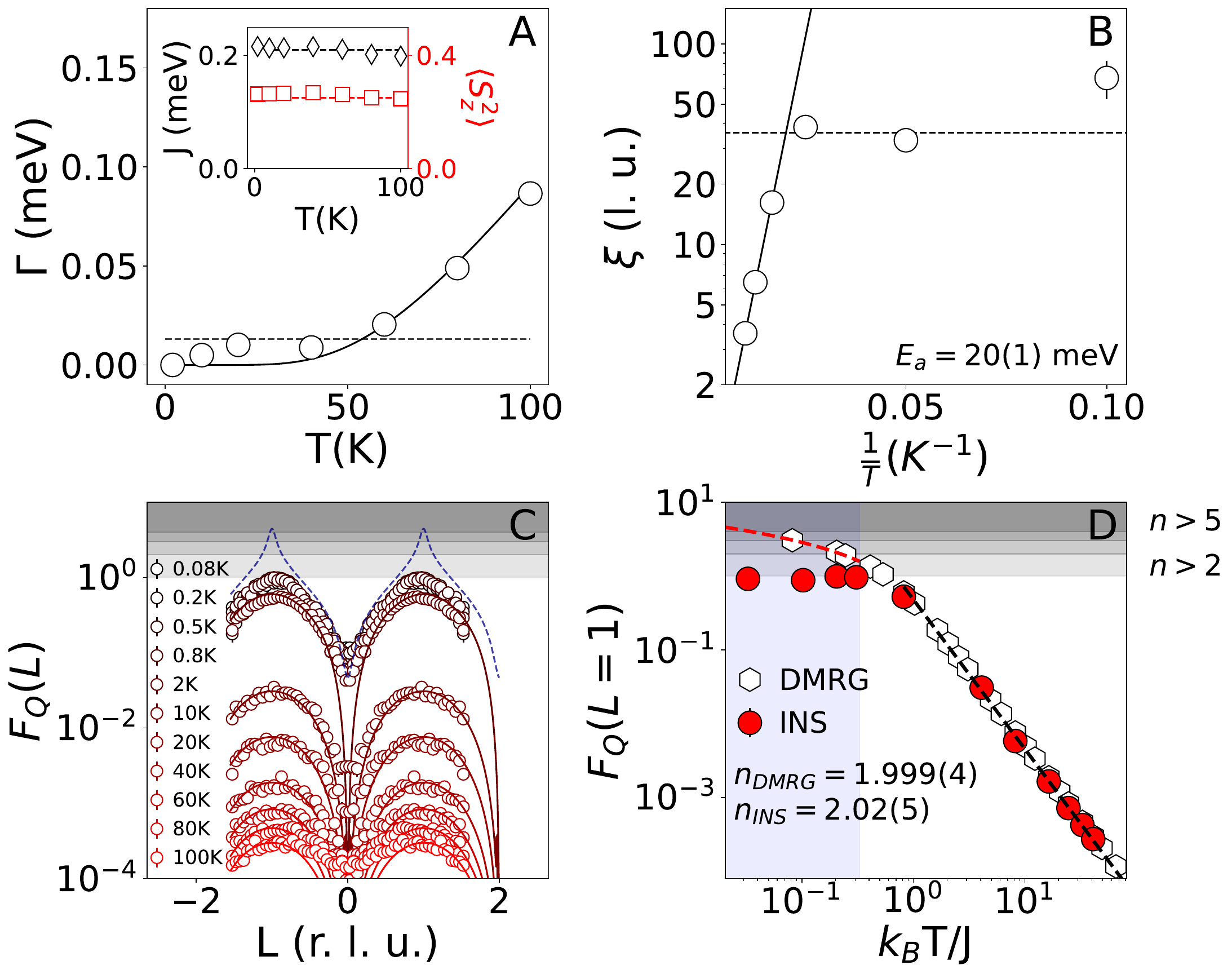}
\caption{{\bf Temperature dependence of INS spectral parameters and quantum Fisher information ($F_Q$).} (A) Life-time broadening parameter as a function of temperature. Dashed line is instrumental resolution HWHM (= 0.013 meV) calculated for $E = 0.5$~meV. Solid curve is a fit to Arrhenius-type exponential function as described in the text. The inset shows fitted exchange interaction, $J$, and integrated intensity, $\langle S_z^2 \rangle$, at different temperatures; horizontal lines indicate nominal values, $J = 0.21$~meV \cite{Wu_2019,Nikitin_PRB2020} and $\langle S_z^2 \rangle = 1/4$. (B) Coherence length calculated using the spinon dispersion and extracted lifetime. Solid and dashed lines are asymptotic Arrhenius and resolution-limited behaviors as in (A). (C) Wave-vector dependence of the QFI, $F_Q(L)$, at various temperatures. Dashed curve is an approximation to asymptotic zero-temperature limit calculated from DMRG data at 200~mK as described in the text. (D) Temperature dependence of maximal quantum Fisher information, $F_Q(L=1)$. Dashed black line is a power-law fit to the data in $T\geq 2$~K range capturing asymptotic high-temperature behavior, $F_Q \sim (J/T)^n$, with $n = 2$. Dashed red curve, shown in the region below $T_N=0.8$~K (shaded), is a fit of DMRG data below 4~K to a logarithmic dependence, $F_Q = \left[ \ln({aJ}/{k_BT}) \right]^\alpha$, with $J=0.21$~meV and fitting parameters $a=1.55(2)$ and $\alpha=1.04(1)$, illustrating the low-T asymptotic behavior; in YbAlO$_{3}$ it is arrested by static order below $T_N$, where part of the excitation spectrum condenses into elastic Bragg peaks that do not contribute to QFI.}
\label{Fig3:t_pars}
\end{figure*}

Figure~\ref{Fig3:t_pars}~(A) shows $\Gamma$ as a function of temperature, revealing no measurable spectral broadening beyond resolution at temperatures below 60~K. Above this point, however, the dispersion does become measurably blurred, with $\Gamma$ eventually reaching an energy-scale of $\sim 0.1$~meV at 100~K, consistent with thermally activated behavior.
An Arrhenius type fit, $\Gamma (T) = \Gamma_0 e^{-\frac{Ea}{k_BT}}$, yields activation energy, $E_a \approx 20$~meV. This energy scale is consistent with thermal population of crystal-field levels other than the ground-state doublet, which invalidates the $S_{\rm eff}=1/2$ description of the Yb ions, leading to information loss.

Using the group velocity of spinons from the dispersion near $L=0$, $v = \pi J/2$, we can obtain an effective spinon coherence length (mean free path), $\xi = v \tau = v \hbar / \Gamma$, shown in Figure~\ref{Fig3:t_pars}~(B) versus the reciprocal of temperature. When measurable broadening does develop at temperatures above 60~K, the effective coherence length appears to track a decreasing exponential trend with increasing temperature.
Fits to an Arrhenius-type model, $\xi=\xi_0 e^{\frac{Ea}{k_BT}}$, where values $\xi > \xi_{0}$ are replaced by a fitted constant $\xi_{0} \approx 35$ in agreement with our resolution limit, are shown over-plotted on the measured data in Figure~\ref{Fig3:t_pars}~(B). According to this analysis, the coherence length exceeds 35 lattice units as it passes beyond the resolution limit of our measurements at staggeringly high temperature of 40~K ($\approx 17 J$). The value of $E_a \approx 20$~meV obtained through this analysis is close to values for the CEF splitting reported in the literature \cite{Wu_2019}. Thus, the major spin-decoherence mechanism at play is likely to be thermal excitation of crystal-field levels outside the $S_{\rm eff}=1/2$ doublet, which presents defects in the chain that are able to change the number of spinons in the system on measurable timescales.

The coherence length encoded in the spinon lifetime ($\xi>$~12~nm) is comparable to the mesoscopic quantum coherence length of Haldane gap magnons observed near zero temperature in spin-1 chains~\cite{Xu_Science2007}. There, however, magnon coherence is quickly lost with the increasing temperature due to collisions that change the quasiparticle content of the excited states and therefore limit the quasiparticle lifetime~\cite{Sachdev_PRL1997, Xu_Science2007, Zheludev_PRL2008}. Consequently, magnons become over-damped at temperatures where thermal energy becomes comparable to the energy of spin interactions. Remarkably, this collisional lifetime mechanism is absent in the case of spinons in the spin-1/2 chain as spinons retain their intrinsic coherence at temperatures much higher than those characteristic of the spin Hamiltonian. 

It is of interest to put our results in the context of quantum metrology, which allows calculating model-independent quantities called entanglement witnesses that can be used to place bounds on the degrees of multipartite quantum entanglement present in the system \cite{Scheie_PRB2021,Scheie_PRB2023Err}. Of specific relevance is the quantum Fisher information (QFI), $F_Q(\hat{A})$ \cite{Hauke_NatPhys2016}, a quantity that can be defined at finite temperature for any system through an imaginary part of dynamical susceptibility with respect to a variable, $\hat{A}$, in that system, $\chi_A''(E)$,
\begin{equation}
    F_Q(\hat{A}) = \frac{1}{4\pi}\int_{0}^{\infty} dE \tanh{ \left( \frac{E}{2k_BT} \right)} \chi_A''(E)
\end{equation}
For the spin-1/2 chain, the QFI, $F_Q$, can be obtained for $\hat{A} = \hat{S}^z$ from the dynamical spin susceptibility at any wave-vector, $\chi''(Q, E)$. Equivalently, it can obtained from the dynamical spin structure factor, $S(Q, E)$, measured by INS (Fig.~\ref{Fig1:Fig1_Exp_Fit_DMRG}), which is related to $\chi''(Q, E)$ via the fluctuation-dissipation theorem \cite{Zaliznyak_PRB1994,Scheie_PRB2021,Scheie_PRB2023Err}.
The obtained QFI can then be used to place lower limits on the level of multipartite entanglement in the system, where QFI $F_Q > n$ at a certain wave vector imply at least $(n+1)$-partite entanglement in the system (Kramers-Rao bound) \cite{Scheie_PRB2021,Scheie_PRB2023Err,Hauke_NatPhys2016} provided no symmetries are taken into account.

Figure~\ref{Fig3:t_pars}~(C) shows the wave-vector-dependent QFI calculated from our neutron spectra for temperatures down to 80 mK (open circles), as well as from our idealized DMRG model (solid curves). The dashed curve represents an approximation to the theoretical maximum at zero temperature, $F_Q|_{T=0} = 4S(Q)$ \cite{Menon_PRB2023}, where $S(Q) = \int_{-\infty}^{\infty} S(Q, E) dE$ is static structure factor given by Fourier transform of the single-time two-point spin correlation function, obtained from DMRG calculations at $200$~mK ($\approx 0.01J/k_B$). The temperature dependence of the maximum quantum Fisher information $F_Q(L=1)$ is shown in Figure~\ref{Fig3:t_pars}~(D) with power-law fits to the asymptotic behavior for both experiment and the idealized DMRG model. Our analysis shows excellent agreement between DMRG and neutron scattering measurements at all temperatures above the magnetic ordering transition, $T_N \approx 0.8$~K. At very low temperatures, $F_Q$ in the idealized model continues to rise, demonstrating at least quadpartite entanglement at $200$~mK. In contrast, in YbAlO$_3$ the QFI is arrested with $F_Q\approx 1$ at $T_N$, though enough spectral weight remains at high energy for it to demonstrate at least bipartite entanglement.

At high-temperature, $F_Q(L=1)$ exhibits a near-perfect $T^{-2}$ power-law decay for both experiment and theory.
Already for temperatures $T \gtrsim 0.5J/k_B$, $F_Q(L=1)$ is below 1, the value where it indicates the presence of at least bipartite entanglement. Thus $F_Q$ as a metric for quantum coherence has limited usefulness at high temperatures. This poses a challenge of developing novel quantum metrology to capture high-temperature quantum behaviors in integrable systems, including the observed coherence of spinon excitations. 

The observed long-range dynamical coherence associated with propagating spinons also contrasts sharply with the local character of single-time two-spin correlation function, $\langle S^z_{j} S^z_{j'} \rangle \approx1/4 \delta_{jj'}$ ($\delta_{jj'}$ is Kronecker delta), at $T \gg J/k_B$ and classical expectation of non-propagative, over-damped or diffusive dynamics in this regime \cite{DeGennes_1958}. Like QFI, the single-time correlation is insensitive to dynamical coherence because it encodes an energy-integrated (single-time) property, static structure factor, $S(Q)$. At high temperature, $S(Q) \approx 1/4$ is $Q$-independent, indicating vanishing single-time spin-spin correlations.

\begin{figure*}[th!]
\includegraphics[width=0.8\textwidth]{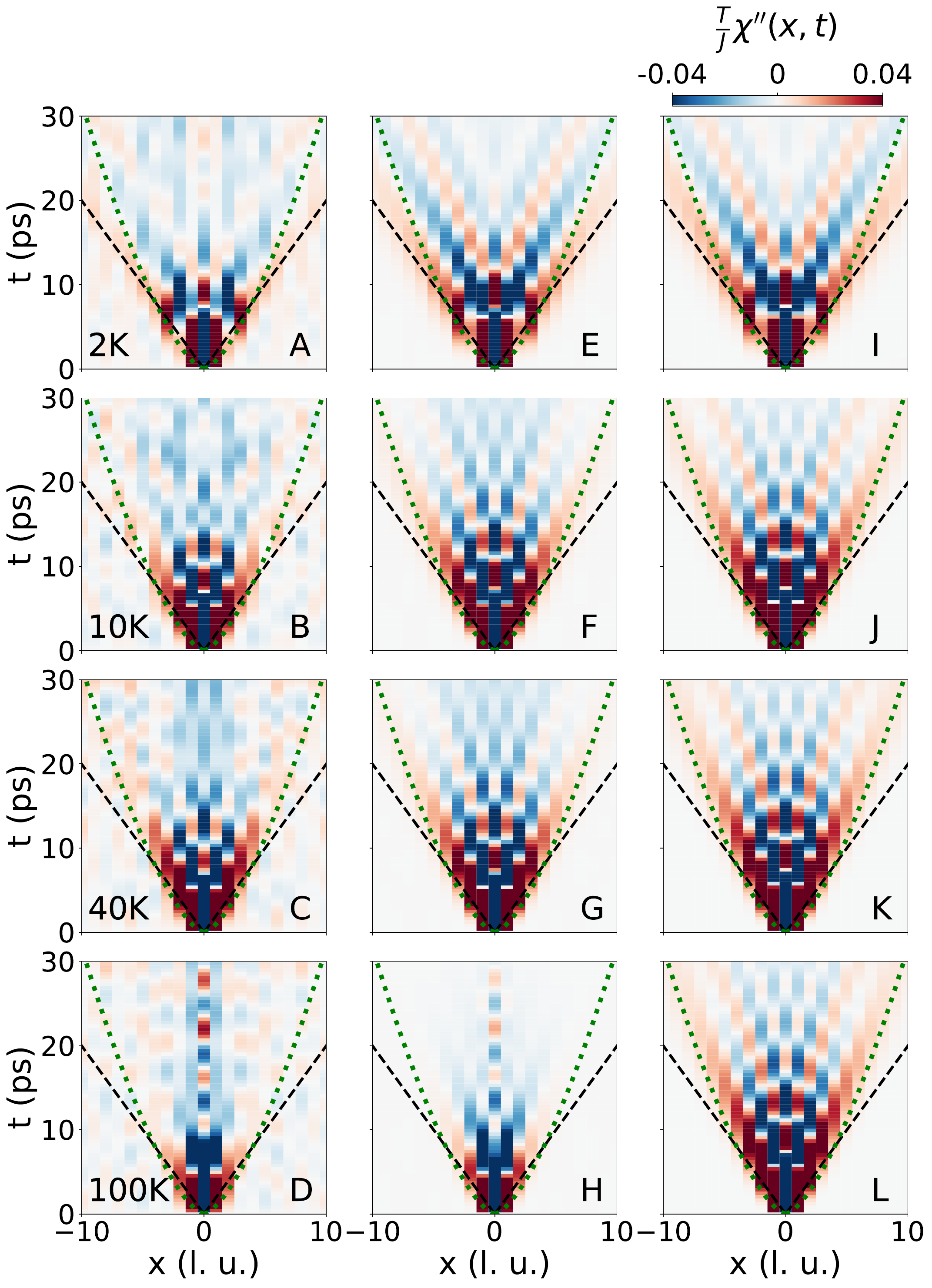}
\caption{{\bf Direct-space and time response functions, $\chi'' (x, t) = -\iu F[\chi'' (Q, E)]$, as a function of temperature.} (A-D) Calculated from inelastic neutron scattering spectra; (E-H) Calculated from fits to inelastic neutron data; (I-L) Obtained from DMRG calculations. Dashed black lines mark the edge of the light-cone in the ballistic regime, $t=\frac{x}{2\pi v}$, while dotted green curves highlight the long-time super-diffusive behavior, $t \sim x^{3/2}$ \cite{Bulchandani2021}, prominent at high temperatures.}
\label{Fig4:fft_2D}
\end{figure*}

The time-dependent, dynamical correlations revealing spinon coherence can be visualized by Fourier transforming the measured $\chi''(Q, E)$ to describe the real-space linear response, $\chi''(x, t)$ \cite{SI,Scheie_NatComm2022}. This is shown in Figure~\ref{Fig4:fft_2D} as a sequence of color-plots scaled by the thermal factor $T/J$, where panels (A-D) show Fourier-transformed (FT) inelastic neutron data, (E-H) show the corresponding Fourier-transforms of our fits to the data, and (I-L) show the space-time theoretical DMRG data. At all temperatures, $\chi''(x, t)$ is measurably nonzero only in a region defined by a coherent ``light cone'' bounded by the spinon velocity and approaching zero width at the origin ($x\rightarrow 0$ as $t\rightarrow 0$) in line with purely local single-time correlations.
This light cone feature corresponds in the wave-vector-energy domain to the dispersive upper boundary of the spinon spectrum, and its presence at high temperatures testifies to the coherent nature of excitations. Remarkably, the linear ballistic transport regime appears to persist on a mesoscopic length scale at short times even when thermal energy scale markedly exceeds interactions.
At long times, however, the transport appears to cross over into a super-diffusive regime, $x \sim t^{2/3}$. Such a super-diffusive behavior is predicted in the high-temperature limit of the Heisenberg chain and has been of interest for some number of years \cite{Bulchandani2021}, but to our knowledge this is the clearest experimental signature of such a behavior to date. At very high temperatures, experiment and fits experience a Lorentzian broadening along the energy axis, which indicates a shortening of the coherence time and a faster decay of dynamical correlations absent in the purely theoretical model, Fig.~\ref{Fig4:fft_2D}~(I-L).

Topologically-protected spinon excitations in integrable systems present an attractive avenue towards encoding information in the spin degree of freedom in materials. Our work demonstrates remarkable quantum coherent behavior of spinons hosted by the effective $S = 1/2$ Heisenberg chains in YbAlO$_3$, including ballistic propagation at temperatures far exceeding the energy-scales at which individual spins interact with each other. The lifetime of these excitations remains longer than our experimental resolution up to very high temperatures, comparable to the crystal-field levels splitting of the Yb$^{3+}$ ions, whose thermal population destabilizes the ground-state Kramers doublets underlying the spin-1/2 Heisenberg chain physics. In turn, this provides a possible control channel for the quantum-collective behaviors in optically-active rare-earth chains where optically-excited ions can be used to control the propagation of information \cite{Awschalom_2018}. Overall, our results suggest that such integrable rare-earth spin-systems may have a far broader range of quantum information applications than previously realized and also challenge quantum metrology to develop new methods suitable for gauging high-temperature dynamical coherence in quantum systems. 

\medskip
\noindent{\bf Methods:}


{\bf{Neutron scattering.}}
The time-of-flight neutron scattering measurements were performed at the Cold Neutron Chopper Spectrometer (CNCS), Spallation Neutron Source (SNS). $E_i = 1.55$~meV ($\lambda = 7.26$~\textrm{\AA}) was used. Here, chopper resolution settings resulted in a resolution full-width-half-maximum (FWHM) of 0.038 meV at the elastic position [see also dashed line in Fig.~\ref{Fig3:t_pars}~(A)]. A single crystal sample of YbAlO$_3$~\cite{Wu_2019} was mounted with its orthorhombic $\ba$ direction vertical, which allowed spectral mapping in the $(0, K, L)$ scattering plane. The wave vector, $Q = (H, K, L)$, is measured in reciprocal lattice units of the orthorhombic $Pbnm$ lattice of YbAlO$_3$ ($a = 5.126$\AA, $b = 5.331$~\AA, and $c = 7.313$~\AA), where Yb-Yb spacing along the chain direction is $d = c/2$. Neutron intensities were binned on a uniform grid in wave-vector and energy, with a focus on two-dimensional slices in the $L-E$ plane. Details of the analyses including fitting to numerical models are described in the Supplementary Information \cite{SI}.

{\bf{Finite-temperature DMRG calculations.}}
The dynamical structure factor (DSF) $S_{\rm DMRG}(q,E)$ vs. one-dimensional in-chain momentum $q \equiv L/2$ and energy $E$ was computed within DMRG in real-time and real-frequency from the retarded correlation function,
\begin{eqnarray}
   S^\mathrm{ret}(x,t)
   &\equiv&
    - \iu \vartheta(t)\, 
      \bigl\langle \hat{S}_x(t) \hat{S}_0^\dagger \bigr\rangle_T
\text{ ,}\label{eq:DSF:xt-1}
\end{eqnarray}
where $\hat{S}_x(t) \equiv e^{\iu \hat{H} t}\hat{S}_x \,e^{-\iu
\hat{H} t}$ is the spin operator $\hat{S}$ acting on site $x$ at
time $t$ in the Heisenberg picture, with $\hat{H}$ the Hamiltonian.
Since DMRG operates on a finite system of length $N=64$ with
open boundary conditions (BCs) and lattice constant $d:=1$, the
DSF was computed relative to the system center, referred to as
origin `0' above, hence having integer $x \in [ -N/2+1, N/2 ]$.
The time evolution was considered up until the light cone was
about to reach the open system boundary. This data was then
zero-padded towards larger system $|x| > N/2$ and extended in
time via linear prediction, followed by double Fourier transform
to momentum $q$ and energy $E$. Additional details are presented
in the Supplementary Information \cite{SI}.

{\bf Acknowledgments}
We are grateful to the SNS staff for invaluable technical assistance and to A. Scheie, C. Broholm, M. Mourigal, and A. Zheludev for valuable discussions. The work at Brookhaven National Laboratory was supported by Office of Basic Energy Sciences (BES), Division of Materials Sciences and Engineering, U.S. Department of Energy (DOE), under contract DE-SC0012704. This research used resources at the Spallation Neutron Source, a DOE Office of Science User Facility operated by Oak Ridge National Laboratory. \\
%
{\bf{Author contributions: }}
I.Z. conceived and directed the study. L.K., I.Z., D.P., A.P. and A.S. carried out neutron scattering experiments and obtained the data. L.K. performed fitting of the neutron spectra. L.K. and I.Z. analyzed the data and prepared the figures. A. W. performed theoretical DMRG calculations. R. K. and A. T. carried out theoretical analyses. L.V. provided the single crystals used in this study. I.Z. and L.K. wrote the paper, with input from all authors.
%
{\bf{Competing Interests: }}
The authors declare that they have no competing interests.
%
{\bf{Data availability: }}
All data needed to evaluate the conclusions in the paper are present in the paper and/or the Supplementary Materials. Additional data available from authors upon reasonable request.


%

\pagebreak
\section*{Supplementary Information}

\hypersetup{pageanchor=false}
\renewcommand{\thepage}{S\arabic{page}}
\setcounter{page}{1}
\renewcommand{\theequation}{S\arabic{equation}}
\setcounter{equation}{0}
\renewcommand{\thefigure}{S\arabic{figure}}
\setcounter{figure}{0} 
\setcounter{table}{0}


\begin{center}
{\bf High-temperature quantum coherence of spinons in a rare-earth spin chain} \\
Lazar L. Kish, Andreas Weichselbaum, Daniel M. Pajerowski, Andrei T. Savici, Andrey Podlesnyak, Leonid Vasylechko, Alexei Tsvelik, Robert Konik, and Igor A. Zaliznyak
correspondence to: zaliznyak@bnl.gov
\end{center}
\bigskip
\noindent{\bf This PDF file includes:}\\
Supplementary Text\\
Supplementary Figures S1-S7\\

\section{Data processing}
\subsection{Inelastic neutron spectroscopy data}\label{SI_subsec:data}
Data was collected by rotating the sample about the vertical direction with the increment of 1 degree within the range of 180$^\circ$ to 360$^\circ$, with the fixed incident neutron energy, $E_i = 1.55$~meV ($\lambda = 7.26$~\textrm{\AA}), resulting in broad-coverage spectral maps over a large region of reciprocal space.

\begin{figure*}[th!]
\includegraphics[width=1.\textwidth]{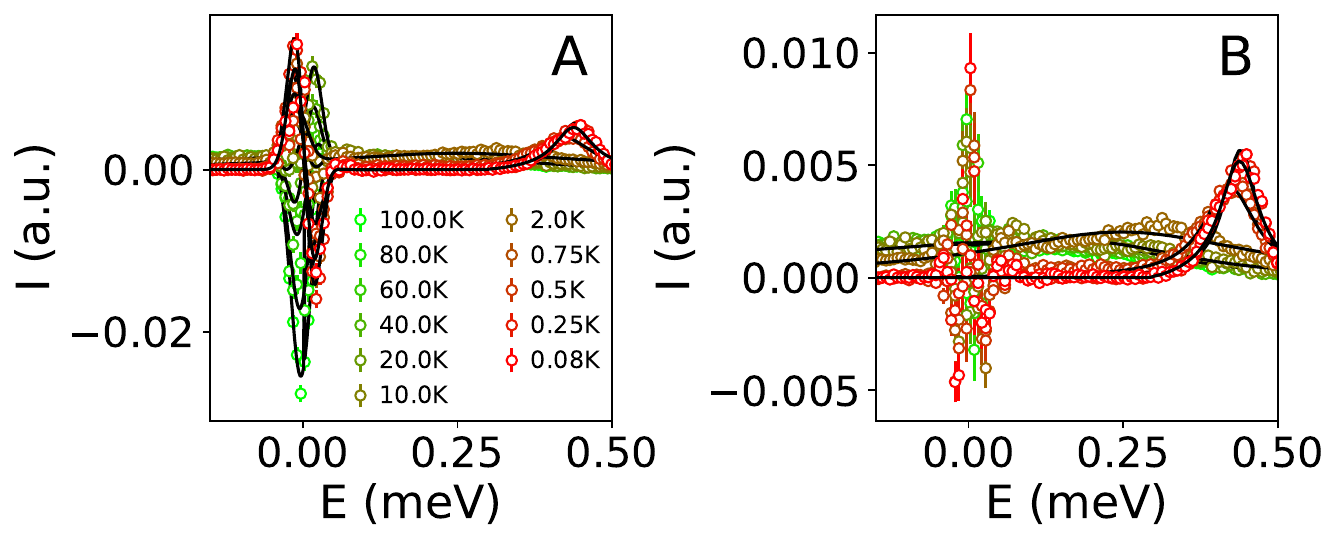}
\caption{{\bf One-dimensional fitting and subtraction of temperature-dependent elastic background.} (A) Data with only high-field BG subtracted (open circles) and fit to a damped harmonic oscillator (DHO) function and a pair of Gaussians (solid line) (B) Data with also the fitted elastic background subtracted off, with DHO fits same as in A (note the different intensity scale).
}
\label{S_Fig1:Qel_bg_sub}
\end{figure*}

Since above the ordering transition, $T_N \approx 0.8$~K, magnetic scattering from YbAlO$_3$ is dispersive only along the chain direction ($c^*$) and only develops a weak dispersion below this temperature, our analyses were carried out on two-dimensional $(L, E)$ slices of $(H, K)$-integrated intensity, with the integration range $H \in [-1.0, 1.0]$ and $K \in [-0.25, 0.25]$). Intensities from a low-temperature high-magnetic-field data set (7~T, 80~mK) were used as the background (BG). Here, the high magnetic field suppresses all inelastic magnetic scatterng from the sample, leaving only structural components from the sample and sample environment. However, this background subtraction procedure leaves a temperature-dependent BG component within the elastic peak region, somewhat over-subtracting the higher temperature data sets [Fig.~\ref{S_Fig1:Qel_bg_sub}]. The residual $Q$-independent BG component is weaker at high-temperature, contrary to expectations for paramagnetic scattering. We believe this component to be nonmagnetic due to its lack of field-dependence. The source of this background may be nuclear scattering from the sample, or temperature-dependent scattering from the sample environment, or perhaps a very slight shift of the instrument's elastic line between different measurements, which introduced a slight systematic bias in our BG subtraction.

In order to isolate this $Q$-independent elastic background component in our zero-field datasets, the intensities were one-dimensionally averaged along the $L$-direction in a region where no Bragg contributions are present at low-temperature, $[0.3, 0.8]$ rlu. The integrated spectrum was fit to a two-component lineshape, where the inelastic magnetic component at all temperatures was well-described by a damped harmonic oscillator response function,
\begin{equation}\label{I_DHO}
I_{DHO}(E)= \frac{A}{1-e^{-E/{k_B T}}} \frac{2 \Gamma E}{(E^2-E_0^2)^2 + 4\Gamma^2 E^2} ,
\end{equation}
with the prefactor, $A$, and damping, $\Gamma$, used as the fit parameters. The residual elastic background component in the zero-field datasets, which shifted slightly in energy between measurements at different temperatures upon subtraction of the high-field data set, was well fit by the superposition of two resolution-limited Gaussian peaks, one fixed to have a positive amplitude ($B_1>0$) and the other fixed to have negative amplitude ($B_2<0$),
\begin{equation}\label{I_BG}
    I(E) = I_{DHO}+I_{BG} = I_{DHO}+B_1 e^{\frac{-(E-E_{c1})^2}{2\sigma_1^2}}+B_2 e^{\frac{-(E-E_{c2})^2}{2\sigma_2^2}} .
\end{equation}

A summary of these fits is presented in Fig.~\ref{S_Fig1:Qel_bg_sub}, where (A) shows high-field-BG subtracted datasets with fits and (B) shows the same data with the fitted residual elastic BG component subtracted.
For every temperature, the peak widths $\sigma_1$ and $\sigma_2$ produced by fits to this function converged to the width of the known resolution function at this energy and instrument configuration ($\sigma_R \approx 0.038/\sqrt{8 \ln 2} \approx 0.016$~meV, where $0.038$~meV is FWHM). These fits to a $Q$-independent elastic background were then subtracted from data sets to provide spectra used for normalization and integration, Fig.~\ref{S_Fig1:Qel_bg_sub} (B).

\subsection{Normalization, weighting, and integration}\label{SI_subsec:data_processing}
Here, we describe the normalization, weighting, and integration procedures performed to obtain structure-factors in absolute units from both experiment and DMRG datasets.

Binned data sets from both experiment and DMRG ($I[Q_i, E_j]$) correspond to a continuous distribution of intensities averaged over each momentum-transfer and energy bin:
\begin{equation}
I[Q_i,E_j] = \frac{1}{\Delta Q \Delta E} \int_{Q_i-\Delta Q/2}^{Q_i+\Delta Q/2}\int_{E_j-\Delta E/2}^{E_j+\Delta E/2} dQ dE I(Q, E) .
\end{equation}

We neglect the Yb$^{3+}$ magnetic form-factor for experimental datasets, which is slowly varying and close to 1 in our measurement range, and set magnetic polarization factor to 1 for moments are nearly orthogonal to our chosen scattering plane. Thus, taking measured intensities to be proportional to the dynamical structure factor, $I(Q,E)=c S^{zz}(Q, E)$, we can obtain normalization using the zero-moment sum rule corresponding to the correlations $S^{zz}$ measured in our dataset,
\begin{equation}
\int_{0}^{1} \int_{-\infty}^{\infty} dQ dE S^{zz}(Q,E) = \frac{1}{4} ,
\end{equation}
\begin{equation}
S^{zz}(Q, E) = \frac{1}{4c} I(Q, E) .
\end{equation}
Our evaluated normalization constant, $c$, is the integral intensity within the first Brillouin zone:
\begin{equation}
c =\frac{1}{2}\int_{0}^{2} \int_{-\infty}^{\infty} dQ dE I(Q,E) \approx \frac{1}{2}\sum_{i, j}I[Q_i,E_j]\Delta Q \Delta E .
\end{equation}
Here, the factor $\frac{1}{2}$ is a result of choosing the conventional crystallographic unit cell for YbAlO$_3$, which contains two spin-chain lattice units and causes the length of the magnetic Brillouin zone to be 2. For our data sets, both in DMRG and experiment, intensity is close to zero outside the region $|E|<0.75$~meV, so the integral was truncated there. From this we calculate $S[Q_i,E_j]$, an average of the continuous function $S(Q,E)$ sampled over the area of each bin.

Instantaneous, single-time and local, single-spin correlations are calculated as integrals of this function over energy and momentum transfer, respectively:

\begin{equation}
S[Q_i]=\sum_j S[Q_i, E_j] \Delta E_j ,
\end{equation}
\begin{equation}
S[E_j]=\sum_i S[Q_i, E_j] \Delta Q_i .
\end{equation}


From the obtained normalized values, $S^{zz}[Q_i, E_j]$, we obtain the imaginary dynamical susceptibility and the integrand for calculation of the quantum Fisher information, $F[Q_i]$, at each $Q_i$:
\begin{equation}
\chi''[Q_i, E_j] = \pi \left(1-\exp{ \left( -\frac{E_j}{k_B T} \right)} \right) S[Q_i, E_j] ,
\end{equation}

\begin{equation}
f[Q_i, E_j] = \frac{4}{\pi} \tanh \left( \frac{E_j}{2k_B T} \right) \chi''[Q_i, E_j] ,
\end{equation}

\begin{equation}
F[Q_i] =\sum_j f[Q_i, E_j]\Delta E_j .
\end{equation}

Because the experimental result includes a convolution with an instrumental resolution function (with an energy FWHM of $\Delta E_{res} \approx 0.038$~meV at $E = 0$ for $E_{i} = 1.55$~meV), the QFI integral was truncated below the elastic line at $E = \Delta E_{res}$ for each measurement.

\subsection{DMRG spectral data}
The dynamical structure factor (DSF) $S_{\rm DMRG}(k,\epsilon)$  vs. momentum $k$ and energy $\epsilon$ was computed within DMRG in real-time and real-frequency from the retarded correlation function, \Eq{DSF:xt-1} in the main text,
\begin{eqnarray}
   S^\mathrm{ret}(x,t)
   &\equiv&
    - \iu \vartheta(t) \underbrace{
      \bigl\langle \hat{S}_x(t) \hat{S}_0^\dagger \bigr\rangle_T
   }_{\equiv\, S(-x, -t)}
\text{ ,}\label{eq:DSF:xt}
\end{eqnarray}
where $\hat{S}_x(t) \equiv e^{\iu H t}\hat{S}_x \,e^{-\iu H t}$ describes the spin operator $\hat{S}$ acting on site $x$ at time $t$ in the Heisenberg picture, with $H$ the Hamiltonian, and $S(x, t) = \bigl\langle \hat{S}_0 \hat{S}_x(t) \bigr\rangle$ is the conventional Van Hove correlation function \cite{DeGennes_1958}. Note that it has different space-time ordering compared to the DMRG correlation function, $S^\mathrm{ret}(x,t)$ in Eq.~\eqref{eq:DSF:xt}, and the two are related assuming space-time homogeneity, $S^\mathrm{ret}(x,t) = \bigl\langle \hat{S}_x(t) \hat{S}_0 \bigr\rangle = \bigl\langle \hat{S}_0 \hat{S}_{-x}(-t) \bigr\rangle = S(-x, -t)$, where dagger is dropped because the spin operator is Hermitian, $\hat{S}_x^\dagger(t) = \hat{S}_x(t)$. Then,
\begin{eqnarray}\label{eq:DSF:S:sym}
  S(-x, -t) &\overset{\eqref{eq:DSF:xt}}{\equiv}&
  \langle \hat{S}_x (t) \hat{S}_0^\dagger \rangle_T
  = \langle \hat{S}_0 (0) \hat{S}_{x}^\dagger(t) \rangle_T^\ast
  = \langle  \hat{S}_0 (0) \hat{S}_{x}(t) \rangle_T^\ast \equiv S^\ast (x,t) ,
\end{eqnarray}
shows that $S(x, t) = S^\ast(-x, -t)$ and therefore $S^\mathrm{ret}(x, t) = - \iu \vartheta(t) S(-x, -t)$ representing Van Hove correlation function for negative times also determines the entire $S(x, t)$.

Since DMRG operates on a finite system of length $N=64$ with open boundary conditions (BCs) and lattice constant $a:=1$, the DSF was computed relative to the system center, referred to as origin `0' above, hence having integer $x \in [ -N/2+1, N/2 ]$.  The time evolution was considered up until the light cone was about to
reach the open system boundary.
Bearing in mind that $S(x,t)$ is zero outside the light cone, this data was zero-padded towards larger system, $|x| > N/2$, prior to Fourier transform to momentum space. The resulting $S(k,t$) was then considerably extended in time by about a factor of 10 via linear prediction, followed by another Fourier transform to frequency, $\omega \equiv \epsilon$ (using $\hbar:=1$). The linear prediction beyond the computed time range permits one to avoid a sharp cutoff in time of the bare DMRG data with its ensuing loss of resolution in frequency space (essentially, allowing a smaller step size in energy). This can be summarized as,
\begin{eqnarray}
   \mathcal{S}(k,\omega) &\equiv&
\int\limits_{-\infty}^{\infty} \tfrac{dt}{2\pi} e^{-\iu\omega t}
    \underbrace{\sum_{x} e^{\iu k x} S(x,t) }_{ \equiv S(k,t) }
 \overset{\eqref{eq:DSF:xt}}{=}
    -\tfrac{1}{\pi} \mathrm{Im}
    \sum_{x} e^{-\iu k x}
    \int\limits_{0}^{\infty} dt\, e^{\iu\omega t}\,
      S^\mathrm{ret}(x,t)
\label{eq:DSF:kw}
\end{eqnarray}
which is the standard formulation of the DSF. It is related to the spectral data of the retarded correlator in the preceding expression, $S^\mathrm{ret}(x,t)$, by relating negative times in $S(x,t)$ to its complex conjugate at positive times while assuming translational invariance, Eq.~\eqref{eq:DSF:S:sym}. This demonstrates that it suffices to compute $S^\mathrm{ret}(x,t)$, or $S(x,t)$ for $t\leq 0$ only.

%

While certain of the above identities, which are exact assuming space-time translational invariance, become approximate in DMRG as it uses finite-size and open BCs, they are used nevertheless. This assumes that as long as the light cone does not reach the open boundary in the real-time evolution, the system is not directly affected by the open boundary. Fully integrating the spectral data in \Eq{DSF:kw} yields the first moment spectral sum rule, $\int\! \frac{dk}{2\pi} \int\! d\omega\ \mathcal{S}(k,\omega) = S(x{=}0,t{=}0) = \langle \hat{S}_0 \hat{S}_0^\dagger \rangle_T = 1/4$ for $\hat{S}_x(t) \equiv \hat{S}^z_x(t)$, which is accurately satisfied in the presented DMRG data.

Upon Fourier transforming the spin operator to momentum space, $\hat{S}_k \equiv \frac{1}{\sqrt{N}} \sum_x e^{\iu k x} \hat{S}_x$, the DSF in \Eq{DSF:kw} can be rewritten as,
\begin{eqnarray}
   \mathcal{S}(k,\omega) &\equiv&
    \int \tfrac{dt}{2\pi}\, e^{\iu\omega t}\,
    \langle \hat{S}_k(t) \hat{S}_k^\dagger \rangle_T
 \ =\ \sum_{ab} \rho_a |\langle a| S_k |b\rangle |^2
     \delta(\omega - E_{ab})
     \ \geq \ 0
\text{ ,}\label{eq:DSF:kw-Lehmann}
\end{eqnarray}
where the second equality shows the Lehmann representation in terms of complete eigenbasis sets $a$ and $b$, having $\hat{H}|a\rangle = E_a |a\rangle$ with $E_{ab} \equiv E_b-E_a$, $\rho_a \equiv e^{-\beta E_a}/Z$ and $Z$ the partition function. As is well known, from Eq.~\eqref{eq:DSF:kw-Lehmann} the detailed balance condition immediately follows, $\mathcal{S}(k, -\omega) = e^{-\beta \omega} \mathcal{S}(k, \omega)$. At zero temperature, the DSF has finite spectral support for $\omega \geq 0$ only, since energy can only be absorbed, but not emitted from the system. By contrast, for large $T\gg J$, 
the DSF $\mathcal{S}(k,\omega)$ becomes symmetric under $\omega \leftrightarrow -\omega$, irrespective of the Hamiltonian, because any transition, $a\to b$, can also be reversed, $b\to a$, with equal probability [cf. \Fig{Fig1:Fig1_Exp_Fit_DMRG}].

\begin{figure*}[th!]
\includegraphics[width=1.\textwidth]{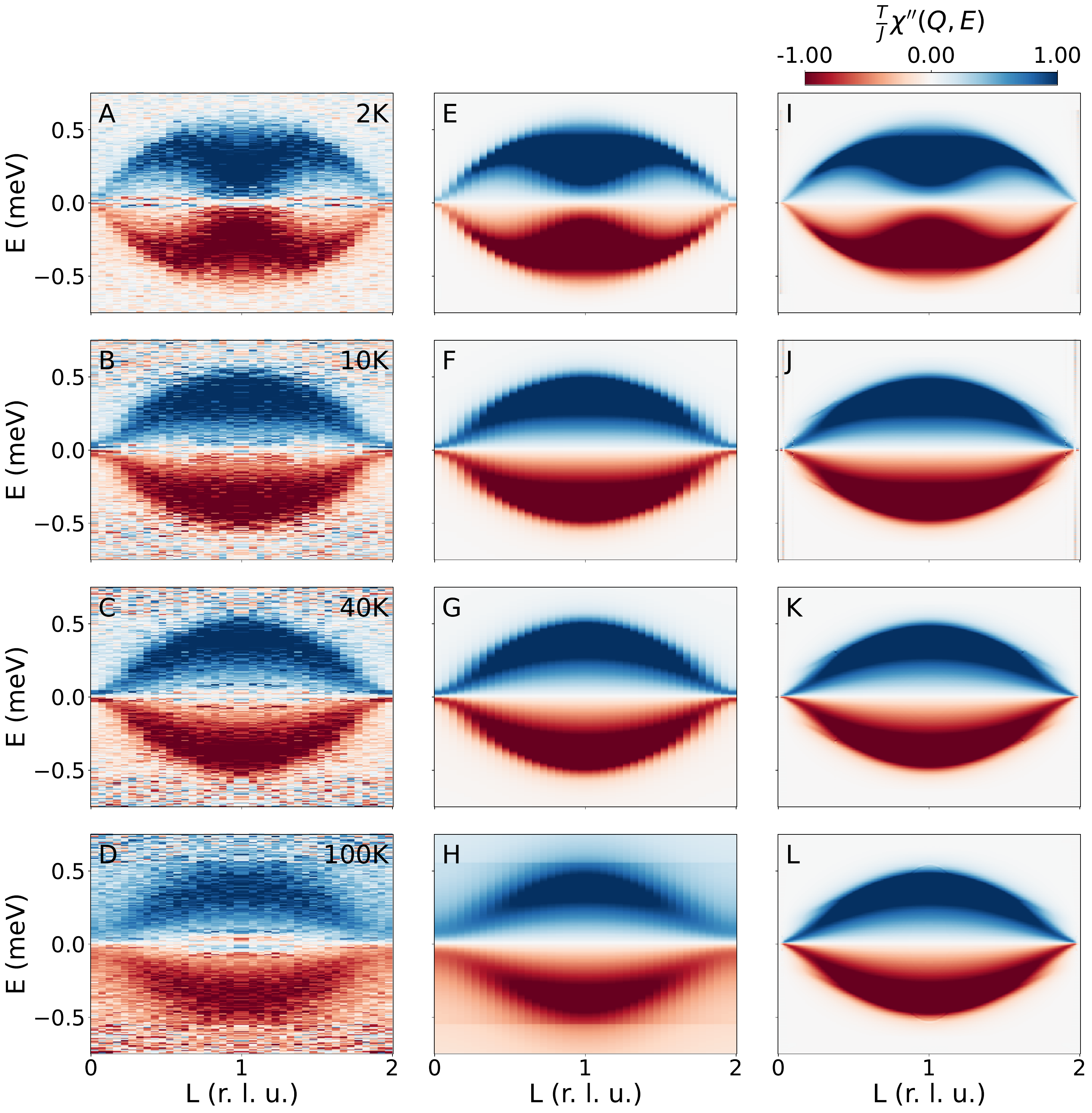}
\caption{{\bf Color plots of $\chi''(Q, E)$ at selected temperatures.} Calculated from (A-D) neutron scattering spectra (E-H) best fits to experimental data (I-L) purely theoretical DMRG calculations.
}
\label{S_Fig2:Chi_2D}
\end{figure*}

\begin{figure*}[th!]
\includegraphics[width=1.\textwidth]{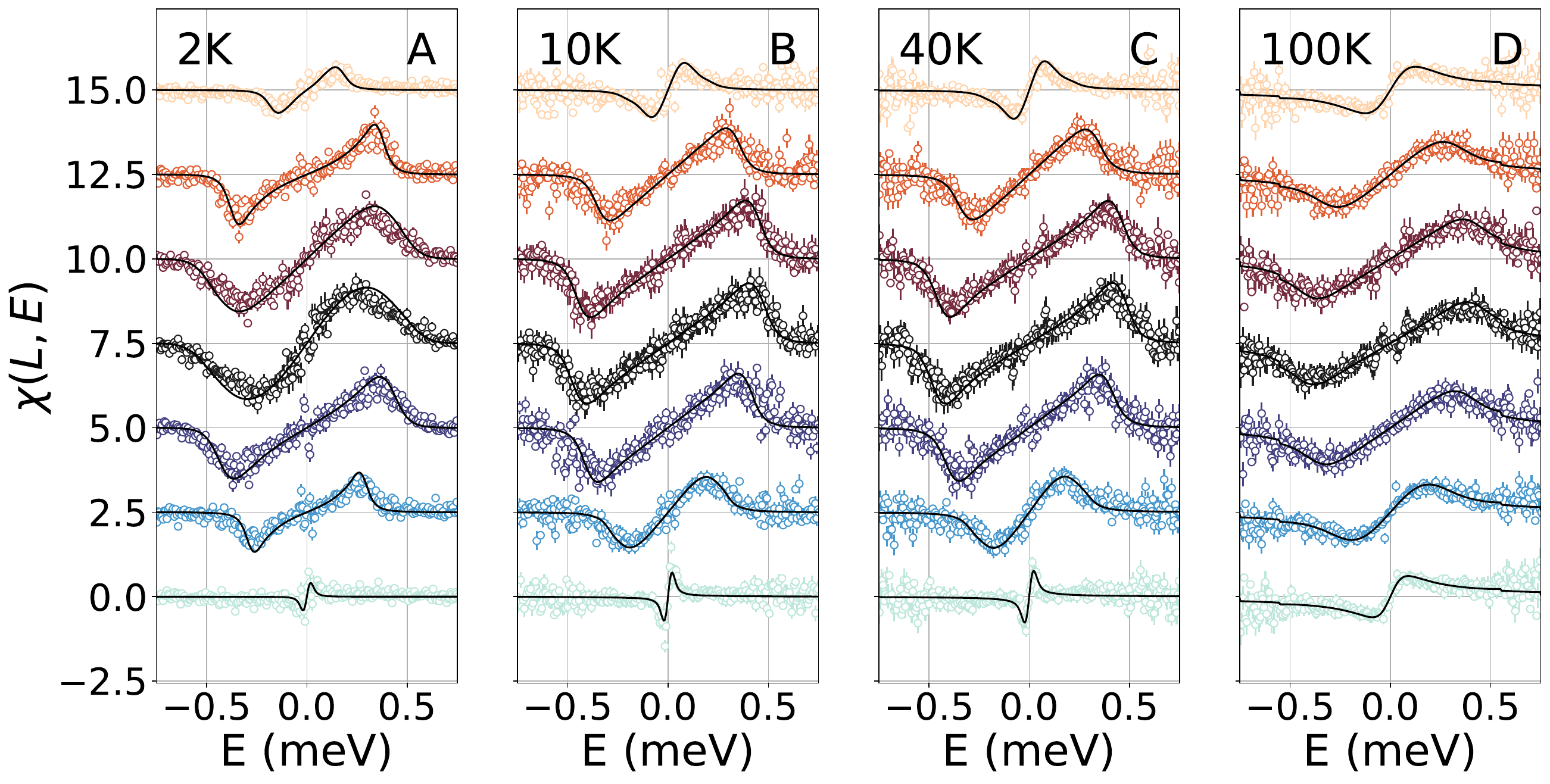}
\caption{{\bf Line cuts of $\chi''(Q, E)$ versus energy at selected temperatures.}
}
\label{S_Fig3:Chi_1D_E}
\end{figure*}

\begin{figure*}[th!]
\includegraphics[width=1.\textwidth]{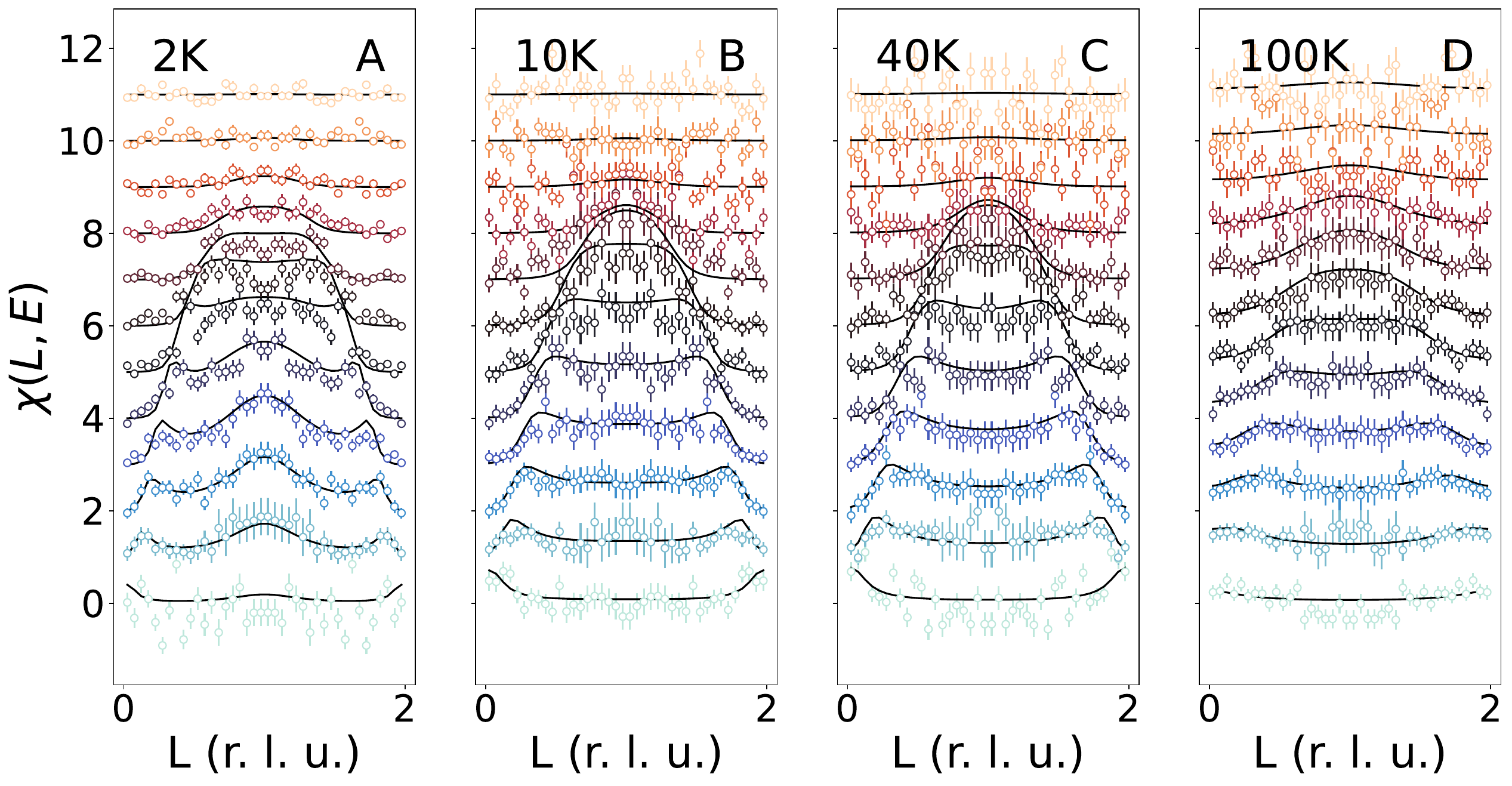}
\caption{{\bf Line cuts of $\chi''(Q, E)$ versus wavevector at selected temperatures.}
}
\label{S_Fig4:Chi_1D_L}
\end{figure*}

\subsection{Fitting to numerical models}\label{SI_subsec:fitting_to_DMRG}
The experimental data were gridded on a uniform grid with wave-vector spacing 0.05 rlu and energy spacing 0.0062 meV to optimize the effective energy resolution. For experimental datasets at each temperature, the corresponding DMRG spectra were fitted to experimental intensities using the Levenberg-Marquart nonlinear least-squares minimization procedure. In each iteration of the fitting procedure, the energy scale (proportional to the exchange-interaction) and intensity prefactor were varied. For each data-point of the model function, the DMRG dataset was resampled by Riemann-integration with a two-dimensional weighting function centered at energy and wavevector coordinates corresponding to the uniformly gridded experimental data:
\begin{equation}
    S_{\rm model}(L_i, E_j) 
    = \frac{\sum_{i', j'} F(L_{i}, k_{i'} \,;\,  E_{j}, \epsilon_{j'}) \,
       S_{\rm DMRG}(k_{i'}, \epsilon_{j'})}{\sum_{i', j'} F(L_{i}, k_{i'} \,;\,  E_{j}, \epsilon_{j'})}
\end{equation}

The result is effectively a convolution of the DMRG predictions with a broadening function $F(L_{i}, k_{i'} \,;\, E_{j},\epsilon_{j'})$, normalized and selected to account for energy-varying instrumental resolution effects as well as the lifetime shortening of magnetic excitations:
\begin{equation}
  F(L_{i}, k_{i'} \,;\, E_{j}, \epsilon_{j'}) 
  = B(L_{i}, k_{i'} \,|\, w, \sigma_L) \            
    V(E_{j}, k_{j'} \,|\, \sigma_E[E_j], \Gamma)  
\end{equation}
This function includes an approximation for the instrumental resolution function on CNCS including binning effects and a constant Lorentzian lifetime-blurring. In our implementation, $F$ is given by a product of two broadening functions $B(L_{i}, k_{i'} \,|\, h, \sigma)$ and  $V(E_{j}, k_{j'} \,|\, \sigma, \Gamma)$  along the respective $k$ and $\epsilon$ directions.

Here, along the wave-vector direction, the function $B$ is the convolution of a Gaussian of width $\sigma_L$ with a window-function of width $h_L$, fixed respectively to be the wave-vector resolution obtained and bin-size used to grid our experimental data along this direction:
\begin{eqnarray*}
    B(L_i, k_{i'} \,|\, h_L, \sigma_L)  
    &=& \tfrac{1}{2} \,\ERF(\tfrac{k_{i'}-L_i+h_L}{2\sigma_L})
      - \tfrac{1}{2} \,\ERF(\tfrac{k_{i'}-L_i-h_L}{2\sigma_L})
    \\ \text{where } \ \ERF(x) &=& \int\limits_{0}^{x} dt \, e^{-t^2}
\end{eqnarray*}
To optimize the energy resolution, we used a relatively large wave-vector bin-size of
$h_L=0.05$ rlu compared to the intrinsic resolution width ($\sigma_L < 0.01$~r. l. u.), which dominated the resolution function along the wave-vector axis.

Meanwhile, the energy broadening was described by the Voigt-function $V$, with an energy-varying Gaussian component width $\sigma_{E_j}$ and constant Lorentzian half-width $\Gamma$. Here, $\sigma_{E_j}$ is given as a function of energy by the resolution function of CNCS as provided by a linear interpolation of reference data provided by PyChop.
\begin{eqnarray*}
  V(E_j, E_{j'} \,|\,  \sigma_{E_j}, \Gamma)   
    &=& \tfrac{\re (w(z))}{\sigma_{E_j} \sqrt{2\pi}} \\
  \text{where } \ z &=& \tfrac{E_j-E_j'+\iu\Gamma}{\sigma_{E_j} \sqrt{2}}\\
  \text{and } \ w(z) &=& e^{-z^2} \ERF(-iz)
\end{eqnarray*}

\subsection{Direct-space and time response function and dynamical susceptibility}\label{SI_subsec:chi''_r_t}
For a system of local magnetic moments, magnetic neutron scattering cross-section measures the dynamical structure factor of the effective spins, which is a Fourier transform (FT) of the two-point spin correlation function,
\begin{equation}\label{S_QE}
    S(Q, E) = \int\limits_{-\infty}^{\infty} \tfrac{dt}{2\pi \hbar} e^{- \frac{\iu}{\hbar} E t}
    \tfrac{1}{N} \sum_{x} e^{ \iu Q x} \langle \hat{S}_0 \hat{S}_x(t) \rangle \, .
\end{equation}
The direct-space and time correlation function, $S(x, t) = \langle \hat{S}_0 \hat{S}_x(t) \rangle$, is obtained from the measured $S(Q, E)$ via inverse Fourier transform,
\begin{equation}\label{S_xt}
    S(x, t) = \int\limits_{-\infty}^{\infty} dE e^{\frac{\iu}{\hbar} E t}
    \sum_{Q} e^{- \iu Q x} S(Q, E) \, .
\end{equation}
While the dynamical spin structure factor is a real function, $S^*(Q, E) = S(Q, E)$, which is clear from Eq.~\eqref{S_QE}, the direct-space and time correlation function is not, $S^*(x, t) = \langle \hat{S}_x(t) \hat{S}_0 \rangle = S(-x, -t) \neq S(x, t)$. This fact is rooted in the absence of time-inversion symmetry, $t \rightarrow -t$, which is broken by the thermal detailed balance condition reflecting the arrow of time. For a system with space-inversion symmetry, $x \rightarrow -x$, such as the spin chains we consider in this work, $S(x, t) = S(-x, t)$. The complex character of $S(x, t)$ somewhat obscures its physical meaning, which recently led to some exotic interpretations of its real and imaginary parts, such as asserting that imaginary part specifically probes quantum nature of spins, distinct from the real part \cite{Scheie_NatComm2022}.

Mathematically, both real and imaginary parts of $S(x, t)$ given by Eq.~\eqref{S_xt} are non-zero because $S(Q, E)$ is neither even nor odd in $E$, which, as mentioned above, is a result of the detailed balance condition, $S(Q, -E) = \exp(-E/k_B T) S(Q, E)$ (here, we consider inversion-symmetric systems, which are invariant with respect to $Q \leftrightarrow -Q$). On the other hand, the imaginary part of the dynamical spin susceptibility, $\chi'' (Q, E)$, contains all the information about dynamical correlations in the system and is odd in energy. Therefore, only the imaginary part of its Fourier transform to direct space and time (given by the sine-Fourier-transform) is non-zero. This direct-space-time response function, $\chi''(x, t) = -\iu F \left[ \chi'' (Q, E) \right]$, contains all of the information about the direct space and time spin correlations in the system and is presented in Figure \ref{Fig4:fft_2D} of the main text.

To understand the relation between the direct-space-time response function, $\chi''(x, t)$, and the spin correlation function, $S(x, t)$, we note that the dynamical spin structure factor is related to the imaginary part of the dynamical spin susceptibility, which is the Fourier transform of $\chi''(x, t)$, via fluctuation-dissipation theorem (FDT) \cite{Zaliznyak_PRB1994},
\begin{equation}\label{Eq_FDT}
    \pi S(Q, E) = \chi''(Q, E) \frac{1}{1 - e^{- \frac{E}{k_B T}}} .
\end{equation}
Taking the (inverse) Fourier transform of both sides of the above Eq.~\eqref{Eq_FDT} we obtain,
\begin{equation}\label{Eq_FDT_FT}
    \pi S(x, t) = \frac{\iu}{2 \pi \hbar} \chi''(x, t) * F\left[ \frac{1}{1 - e^{- \frac{E}{k_B T}}} \right] ,
\end{equation}
where we have used the property that the Fourier transform of a product of two functions is a convolution of the Fourier transforms of these functions [prefactor $1/2 \pi \hbar$ follows from the definition of FT in Eqs.~\eqref{S_QE}, \eqref{S_xt}]. The Fourier transform of the detailed balance factor is straightforwardly evaluated,
\begin{equation}\label{Eq_DBF_FT}
   \frac{\iu}{\pi \hbar} F \left[ \frac{1}{1 - e^{- \frac{E}{k_B T}}} \right] = \iu \delta(t) - \frac{1}{\hbar} k_B T \coth(\pi k_BT t /\hbar) ,
\end{equation}
which shows that the imaginary part of $S(x, t)$ is, up to a $2\pi$ multiplier, simply the direct-space-time response function, $2 \pi \mbox{\rm Im} \left[ S(x, t) \right] = \chi''(x, t)$, while its real part is given by a convolution of the same $\chi''(x, t)$ with the system-independent function, $\coth(\pi k_BT t /\hbar)$.

Another, perhaps simpler way to reach the same conclusion is to decompose the dynamical spin structure factor, $S(Q, E)$ [Eq.~\eqref{S_QE}], into an $E$-odd and $E$-even parts,
\begin{equation}\label{S_QE_odd_even}
    S(Q, E) = \frac{1}{2} \left( S(Q, E) - S(Q, -E) \right) + \frac{1}{2} \left( S(Q, E) + S(Q, -E) \right) \, .
\end{equation}
We then notice that by virtue of the FDT, Eq.~\eqref{Eq_FDT}, and $\chi'' (Q, E) = -\chi'' (Q, -E)$, which holds for systems with spacial inversion symmetry such as spin chain we consider, the odd part is just
\begin{equation}\label{S_QE_odd}
   S_{odd}(Q, E) = \frac{1}{2} \left( S(Q, E) - S(Q, -E) \right) = \frac{1}{2 \pi} \chi'' (Q, E) \, ,
\end{equation}
while the even part is,
\begin{equation}\label{S_QE_even}
   S_{even}(Q, E) = \frac{1}{2} \left( S(Q, E) + S(Q, -E) \right) = \frac{1}{2 \pi} \coth \left( \frac{E}{2 k_B T} \right) \chi'' (Q, E) \, .
\end{equation}
Hence, as before, we obtain,
\begin{align}\label{S_xt_odd_even}
    \mbox{\rm Im} \left[ S(x, t) \right] & = \frac{1}{2 \pi } \chi''(x, t), \\
    \mbox{\rm Re} \left[ S(x, t) \right] & = - \frac{1}{2 \pi } \chi''(x, t) * \coth(\pi k_BT t /\hbar) \, .
\end{align}
This, once again, demonstrates that all information about dynamical spin correlations in the system is contained in the direct-space-time response function, $\chi''(x, t)$, or, equivalently, in the imaginary part of the direct-space-time correlation function, $S(x, t)$.

\subsection{Calculation of direct-space and time correlation functions}\label{SI_subsec:FT_r_t}
Calculations of direct-space and time-domain dynamical spin susceptibility (response function), $\tilde{\chi}''(x, t)$, were carried out using each of the neutron scattering, fit model, and DMRG datasets. This quantity is the two-dimensional inverse fast-Fourier transform (iFFT) of the imaginary part of the dynamical spin susceptibility which is calculated directly from the dynamical structure factor via Eq.~\eqref{Eq_FDT},
\begin{equation}\label{FDT}
    \chi''(Q, E) = \pi \left( 1-e^{- \frac{E}{k_B T}} \right) S(Q, E)
\end{equation}
This two-dimensional iFFT defined on a regularly gridded dataset is given by the summation:
\begin{equation}
    \chi''[x_j, t_k] = \sum_{n, m}^{N, M} e^{2 \pi i \frac{x_j Q_n}{N}} e^{2 \pi i \frac{t_k E_m}{M}} \chi''[Q_n, E_m]
\end{equation}
Here, $n, m$ index the wave-vector and energy points, while indices $j$ and $k$ are new position and time indices, and the summation is taken over all measured energies and wavevectors within the first Brillouin zone. The finite energy and wave-vector bin-size of gridded data leads to a cutoff at large time and positions while restriction of the sum to the first Brillouin zone leads to a quantization of the position axis with a spacing of 1 lattice unit. As measured intensities were approximately zero at high energy transfers above $\Delta_E = 0.75$~meV, intensity points were padded with zeros down to -5~meV and up to 5~meV to increase the energy range and thus allow the sampling resolution of the calculation in the time-domain to be comparable to the DMRG calculation.

As $\chi''(Q, E)$  is an odd function in energy, its Fourier transform is completely imaginary. Moreover, this suppresses elastic components which otherwise carry a significant background in our experimental datasets and present a major issue for the Fourier-transform. In order for the results to be directly comparable, the experimental and fitted datasets were treated on an equal footing, undergoing the same sequence of operations during data processing. Data were anti-symmetrized along the energy axis and and symmetrized along the wave-vector axis such that $\chi''[Q_n, -E_m]=-\chi''[Q_n, E_m]$ and $\chi''[-Q_n, E_m]=\chi''[Q_n, E_m]$. This was necessary due to the partial coverage of the first Brillouin zone obtained during our measurement. For the data set at 2~K only, data were excluded from the symmetrization below an energy transfer of $\Delta E = -\frac{k_BT}{2}=0.086$~meV to avoid amplifying noise from the negative energy side where the signal is weak due to the thermal balance factor.

Figure~\ref{S_Fig2:Chi_2D} shows color plots of the calculated $\chi''[Q_n, E_m]$ for selected temperatures in each of our datasets, while Figure~\ref{S_Fig3:Chi_1D_E} and~Figure\ref{S_Fig4:Chi_1D_L}, respectively, show line cuts of the resulting $\chi''[Q_n, E_m]$ at selected temperatures. The complex inverse fast-Fourier transforms were evaluated from these datasets using the standard Cooley-Tukey algorithm.

All neutron spectroscopy data are intrinsically filtered by an instrumental resolution function in the 4-dimensional momentum-Energy domain, which leads to an extra time-dependent decay factor in the Fourier-transform of spectroscopic data. This instrumental effect can be corrected for by assuming that the measured susceptibility function $\chi''_{meas}(k, E)$ is approximated by a convolution of the intrinsic sample susceptibility function with a normalized Gaussian profile $G(\sigma_R, E)=\frac{1}{\sigma\sqrt{2\pi}}\exp(-\frac{x^2}{2\sigma_R^2})$ of standard deviation $\sigma_R$ determined by the instrument resolution:
\begin{equation}
\chi''_{R}(k, E) = \chi''(k, E) * G(\sigma_R, E)
\end{equation}
Such a convolution along the energy axis meanwhile corresponds to a multiplication of the Fourier-transformed intrinsic $\chi''(x, t)$ with the Fourier-transform of the resolution function in the time-domain:
\begin{equation}
   \chi''_{meas}(x, t)= F(\chi''(k, E) * G(\sigma_R, E)) = \chi''(x, t)g(s, t)
\end{equation}
Here, the Fourier-transformed resolution function $g(s, t)=\exp(-\frac{t^2}{2s^2})$ is also Gaussian with standard deviation $s=\frac{1}{2\pi \sigma_R}$ and normalized to $g(s, t=0)=1$.

In our calculations of $\chi''(x, t)$ shown in the main text, the energy resolution FWHM was assumed to be constant and equal to the value at the elastic line, $\Delta E_{res}|_{E = 0} = 0.038$~meV ($\sigma_R=0.016$~meV).  The results shown in Fig.~\ref{Fig4:fft_2D} are therefore the Fourier-transform of measured intensities, corrected by this approximation to the Fourier-transformed resolution:
\begin{equation}\label{chi_Res_corr}
    \chi''(x, t) = \frac{\chi''_{meas}(x, t)}{g(\frac{1}{2\pi \sigma_R|_{E=0}}, t)} .
\end{equation}

\subsection{Jordan-Wigner fermion analysis of the two-spinon structure factor at high temperature}\label{SI_subsec:JW_fermions}
Here, we present an analysis of our data using a theoretically motivated semi-phenomenological spectral function of a spin-1/2 Heisenberg chain obtained by using Jordan-Wigner (JW) fermionization. This analysis, which we used in our initial approach to the problem, leads to very similar results and conclusions as obtained from the numerically exact analysis using the full DMRG calculation described in the main text, thus providing an additional support for our findings.

We begin by considering a chain of spins S = 1/2 with an XXZ Hamiltonian,
\begin{eqnarray}\label{XXZ_model}
H = \sum_r \Big[ J_{} (\s^x_r\s^x_{r+1} + \s^y_r\s^y_{r+1}) + J_z\s^z_r\s^z_{r+1}  \Big] ,
\end{eqnarray}
where $\s^\alpha_r$ ($\alpha = x, y, z$) are spin-1/2 operators and $r$ numbers sites ($1,...,N$) of a 1D lattice with spacing $a = 1$. Using the Jordan-Wigner transformation, this model Hamiltonian can be recast into a fermionic form \cite{Tsvelik_book2003},
\begin{eqnarray}\label{XXZ_JW_fermions_r}
H = \sum_r \Big[ \frac{J_{}}{2} \left( c^+_{r}c_{r+1} + H.c. \right) + J_z \left( c^+_{r}c_{r} - 1/2 \right) \left( c^+_{r+1}c_{r+1} - 1/2 \right) \Big] ,
\end{eqnarray}
or, upon lattice Fourier transform, $c_{q} =  \frac{1}{\sqrt{N}} \sum_r e^{-\iu q r} c_{r}$, $c_{r} = \frac{1}{\sqrt{N}} \sum_q e^{\iu q r} c_{q}$,
\begin{eqnarray}\label{XXZ_JW_fermions_q}
H = \sum_q \Big[ \left( J_{}\cos{q} - J_z \right) c^+_{q}c_{q} + J_z \sum_{q',k} \cos{k}\, c^+_{q+k}c^+_{q'-k} c_{q'}c_{q} \Big] .
\end{eqnarray}
The absence of an average ordered spin imposes the half-filling condition of the JW fermion band, $\frac{1}{N} \sum_{q} \langle c^+_{q}c_{q} \rangle = \frac{1}{N} \sum_r \langle c^+_{r}c_{r} \rangle = \frac{1}{2} - \frac{1}{N} \sum_r \langle \sigma_r \rangle = \frac{1}{2}$, which is enforced by the interaction term.

For $J_z = 0$, in the XY case, the interaction term is not present and the model is reduced to that of free fermions, which is solved exactly. Here, we are interested in the limit $T>> J_{}, J_z$.
At $J_z=0$, the spin susceptibility, $\chi_{zz}$, is given by the polarization loop:
\begin{eqnarray}\label{JW_fermion_chi}
\chi_{zz}(\omega,q) = \frac{1}{(2\pi)}\int {\mbox{d}} k \frac{n(\epsilon_k) - n(\epsilon_{k+q})}{\omega + \iu 0 - \epsilon_{k} + \epsilon_{k+q}}.
\end{eqnarray}
At high temperatures, $n(\epsilon) \approx 1/2 - \epsilon/4T$. Substituting this into the previous formula we obtain,
\begin{eqnarray} \label{Chi_zz_XY}
\chi_{zz}(\omega,q) = \frac{1}{8T} \Big[ 1 - \frac{1}{\sqrt{1 - \Big(\frac{2J_{}\sin q/2}{\omega +\iu 0}\Big)^2}}\Big].
\end{eqnarray}
The susceptibility is real at high (absolute) frequencies, beyond $\omega^2 > ({2J_{}\sin q/2})^2$, and at zero frequency where it follows the Curie law. The non-zero imaginary part of the dynamical spin susceptibility corresponds to a continuum within the two-particle boundary, $|\omega| \leq {2J_{}\sin q/2}$,
\begin{eqnarray} \label{Chi''_zz_XY}
\chi''_{zz}(\omega,q) = \frac{\omega}{8T} \frac{\theta \left( ({2J_{}\sin q/2})^2 - \omega^2 \right)}{\sqrt{({2J_{}\sin q/2})^2 - \omega^2}}.
\end{eqnarray}

The spectral weight of $\chi''_{zz}(\omega,q)$ in Eq.~\eqref{Chi''_zz_XY} diverges at the upper boundary of the continuum, $|\epsilon_{u}(q)| = 2J_{}\sin q/2$, similarly to the zero temperature case. An account for the interaction term in Eqs.~\eqref{XXZ_JW_fermions_r}, \eqref{XXZ_JW_fermions_q}, removes this divergence in the isotropic XXX chain, $J_z = J$. Instead, at $T = 0$ there is a similar singularity at the lower boundary of the continuum. This divergence, however, as well as the sharp lower boundary itself, result from a step-like Fermi distribution function of the JW fermions at $T = 0$ and should be expected to smear away with the increasing temperature.

The interaction term in Eqs.~\eqref{XXZ_JW_fermions_r}, \eqref{XXZ_JW_fermions_q} can be treated as a perturbation, or using a mean field decoupling in a random phase approximation. The main temperature-independent effect of the interaction is to renormalize the fermion dispersion by a factor $\pi/2$, such that the upper boundary of the two-particle continuum becomes, $|\epsilon_{u}(q)|  = \pi J_{} \sin q/2$. At finite temperatures, the perturbation theory in $J_z$ is dominated by the real part of $\chi_{zz}$. The calculation of the Feynman  diagrams shows that they contain singularities at the upper threshold of the continuum, which is consistent with the removal of the upper-boundary singularity in the isotropic XXX case.

In the high-temperature regime where all fermion states are equally populated, $n(\epsilon_q) \approx 1/2$, the spectral distribution of the dynamical structure factor, $S (q, \omega) = \chi''_{zz}(\omega,q) \left[ \pi (1 - e^{-\omega/T}) \right]^{-1} \approx \left( T/ \pi \omega \right) \chi''_{zz}(\omega,q)$ can be expected to become uniform in energy, $S(q, \omega) \equiv S(q)/ \left[ \omega_{+}(q) - \omega_{-}(q) \right]$, within the continuum boundaries, $\omega_{+}(q)$ and $\omega_{-}(q)$, defining its support in $(q, \omega)$ space. At the same time, the energy-integrated dynamical structure factor, which describes static correlations, becomes $q$-independent, $S(q) = 1/4$, reflecting the vanishing correlation length at $T \gg J$. Hence, we phenomenologically modify the imaginary dynamical susceptibility of Eq.~\eqref{Chi''_zz_XY} to describe the measured spinon continuum as,
\begin{eqnarray} \label{Chi''_zz_XXX}
\chi''_{zz}(\omega,q) = A \frac{\omega}{T} \frac{\theta \left( 2 J_{} \sin q/2 - \omega \right)\theta \left( 2 J_{} \sin q/2 + \omega \right)}{2 J_{} \sin q/2} .
\end{eqnarray}
Here, the denominator makes $S(q)$ $q$-independent, corresponding to zero correlation length, and the prefactor $A$ ensures that the integral spectral weight satisfies the first moment sum rule.

Eq.~\eqref{Chi''_zz_XXX} uses the free fermion dispersion of the upper boundary, $\epsilon_{u}(q) = 2J_{}\sin q/2$. Hence, the effective exchange interaction obtained by fitting the lowest-temperature data can be expected to be renormalized as $\tilde{J} = \pi/2J$, to account for the interaction-renormalized spinon dispersion in the XXX case, $J_z = J$, compared to the free-fermion XY case, $J_z = 0$.

\subsection{Fitting to the semi-phenomenological step-function fermion model}\label{SI_subsec:stepfnfit}
By fitting our experimental data to the numerically precise theoretical DMRG results, the analysis presented in the main text quantifies blurring of the spectrum and hence the finite lifetime and decoherence of spinons due to environmental factors external to the quantum spin-1/2 Hamiltonian. Here, we fit our data to the semi-phenomenological fermion model for the two-spinon spectrum given by Eq.~\eqref{Chi''_zz_XXX}, where $\chi''(q, E)$ is zero outside the upper two-spinon boundary (before blurring) and linear in energy below it. This analysis incorporates blurring of the continuum boundary due to factors intrinsic to the Hamiltonian, such as multispinon excitations, into the same phenomenological blurring parameter, $\gamma$, as the extrinsic finite lifetime decoherence, thus presenting the lower boundary for the latter. By comparing with the analysis in the main text, this also allows to gauge the relative importance of the two effects.

\begin{figure*}[t!h!]
\includegraphics[width=0.7\textwidth]{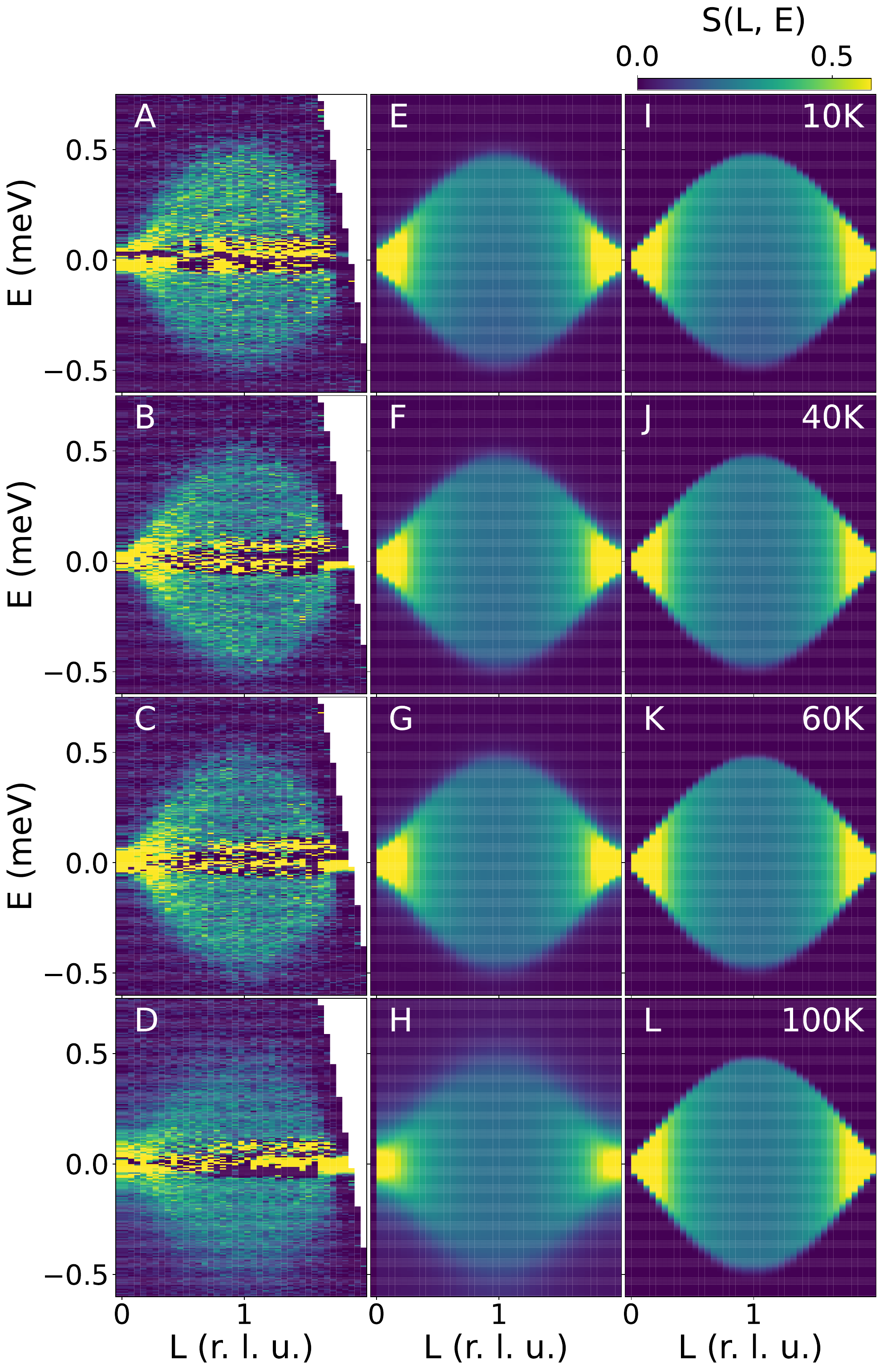}
\caption{{\bf The spinon spectra in YbAlO$_{3}$ at different temperatures, fit to the phenomenological step-function model for the high-temperature behavior. } (A-D) Color contour maps of the spectral density of the measured neutron scattering intensity at different temperatures. These data are integrated in the dispersionless transverse directions with $K=[-1.0, 1.0]$ and $H=[-0.25, 0.25]$. (E-H) Fits to phenomenological model with numerical broadening accounting for energy-varying resolution and spinon lifetime as described in the Supplemental text, and directly comparable to neutron data. (I-L) Phenomenological step-function model including only resolution blurring to describe the spectrum without lifetime effects.}
\label{S_Fig5:Stepfn_fit}
\end{figure*}

We use the dynamical structure factor, $S_{\rm f} (q, E)$, for the unblurred model of Eq.~\eqref{Chi''_zz_XXX} given by:
\begin{equation}
S_{\rm f}(q, E) = \mathop{A} \mathop{\frac{E}{k_BT}} \frac{1 + \mathop{\mathrm{sgn}} \left( \epsilon^2(q)-\omega^2 \right)}{2\epsilon{(q)} \left( 1-e^{-E/k_B T} \right)}
\end{equation}
Here, $\epsilon(q) = \pi J\sin{(\pi L)}$ describes the dispersion of the boundary and ${\mathrm{sgn}}$ in the numerator implements the step function, yielding zero intensity for $|E| > |\epsilon(q)|$. The denominator includes a normalization by $\epsilon(q)$ accounting for our observation that the energy-integrated structure factor is constant at high temperature, and the thermal detailed balance factor converts from $\chi''(q, E)$ to $S(q, E)$. In order to account for broadening effects, this model was numerically convoluted with the broadening function, $F$, as described in Supplemental section~\ref{SI_subsec:fitting_to_DMRG},
\begin{equation}
    S_{\rm model}(L_i, E_j) 
    = \frac{\sum_{i', j'} F(L_{i}, k_{i'} \,;\,  E_{j}, \epsilon_{j'}) \,
       S_{\rm \theta}(k_{i'}, \epsilon_{j'})}{\sum_{i', j'} F(L_{i}, k_{i'} \,;\,  E_{j}, \epsilon_{j'})}
\end{equation}
Again, the broadening function $F$ includes the effects of energy-varying resolution broadening and a Lorentzian broadening parameter which describes limitations of the spinon lifetime, and in the present case also the blurring of the upper continuum boundary due to multi-spinon excitations. The $J$, amplitude, and Lorentzian broadening parameters were allowed to vary. Supplemental Figure~\ref{S_Fig5:Stepfn_fit} shows results of this fitting, which are analogous to Figure~\ref{Fig1:Fig1_Exp_Fit_DMRG} of the main text, with panels A-D showing neutron scattering data, E-H showing fits to the model, and I-L showing the model without Lorentzian blurring.

The line-cuts of the fitted model and the experimental data along the energy axis presented in Supplemental Fig.~\ref{S_Fig6:waterfall_fit_stepfn} show the good agreement between the model and the data. The main qualitative features of the experimental and DMRG data at high-temperature are reasonably well reproduced by this model. However, it shows less quantitative agreement with the data as demonstrated by the reduced $\chi^2$ parameter increasing from $r. \chi^2\sim 2$ to $r. \chi^2 \sim 5$ in going from the first-principles DMRG analysis to the phenomenological model (see figure captions).

The temperature dependence of model parameters is presented in Supplemental Fig.~\ref{S_Fig7:t_pars_stepfnmodel}, which is analogous to Figure~\ref{Fig3:t_pars} of the main text, but now using the step-function model to describe the temperature dependence of the effective spinon lifetime. Here, we see an additional degree of broadening above resolution, which is due to the multi-spinon states. This effect appears to be rather small, reducing the estimate for effective spinon coherence length-scale to $\xi_{step} = 13.4$~l. u.

It is important to note that where the DMRG analysis in the main text examines how the coherence length of quasiparticle excitations is limited by coupling of the spin-chain subsystem to an external heat-bath, the step-function analysis instead measures how the two-spinon upper boundary is affected by multispinon excitations as well as external factors. The effect of multispinon-states is barely observable in our measurements, yielding an energy broadening similar to the width of the instrumental resolution evaluated at $E=0.5$~meV, and smaller than the instrumental resolution for the rest of the measured energy range. The temperature dependence obtained from this analysis correlates well with our earlier observations, showing Arrhenius-like behavior with a gap energy of $E_a=19$~meV similar to the DMRG fitting.

\begin{figure*}[t!h!]
\includegraphics[width=0.8\textwidth]{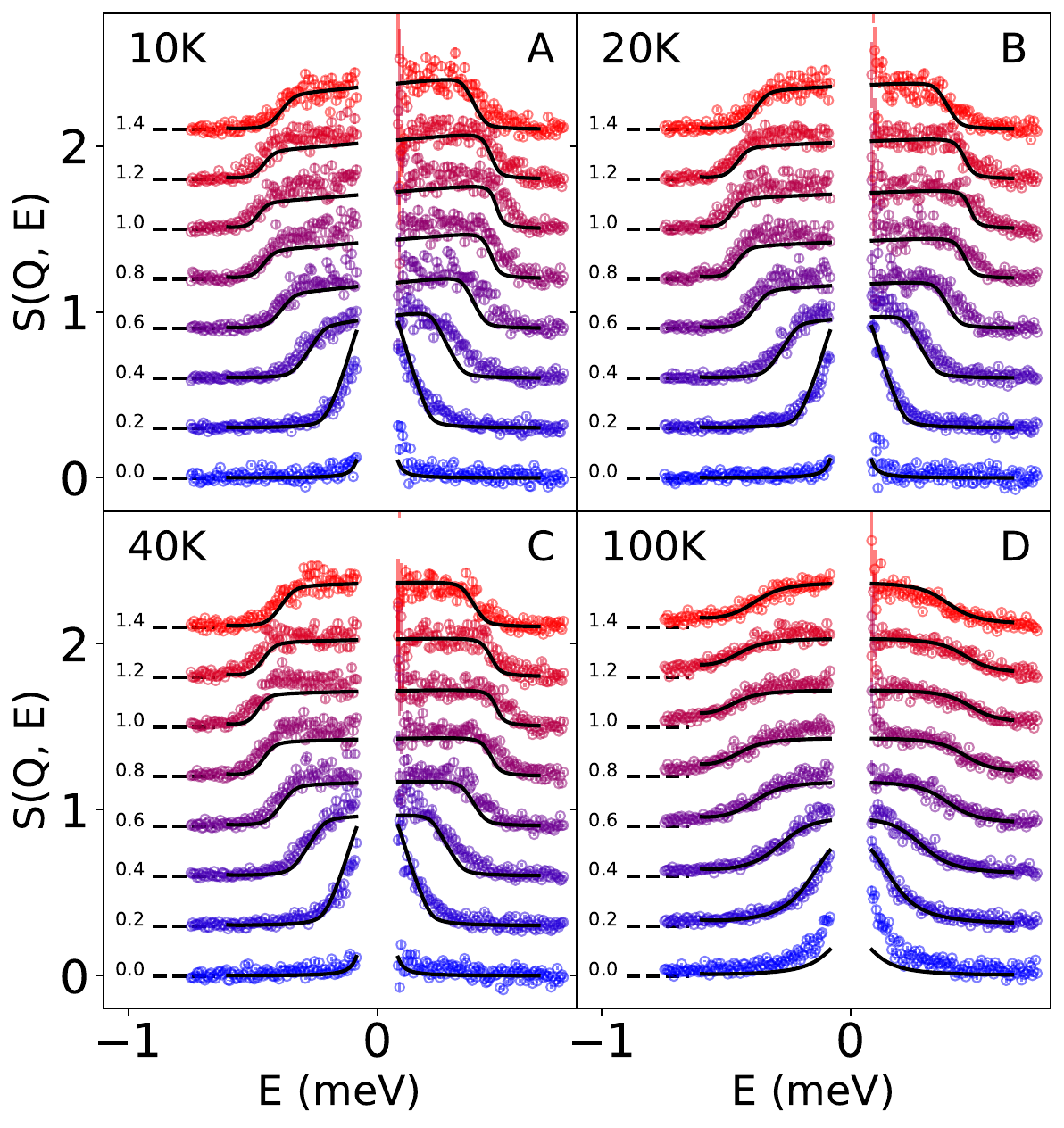}
\caption{{\bf Line cuts along the energy axis of our data and fits to the phenomenological step-function model. } Curves are given an incremental offset for visualization, with dashed leader-lines from each curve signifying the zero of intensity. The labels next to each curve signify the central L value of each line-cut, which are 0.2 r. l. u. wide. (A)~10~K~(r.~$\chi^2 = 6.5$); (B)~20~K~(r.~$\chi^2 = 5.8$3); (C)~40~K~(r.~$\chi^2 = 5.7$); (D)~100~K (r.~$\chi^2 = 5.04$)}
\label{S_Fig6:waterfall_fit_stepfn}
\end{figure*}

\begin{figure*}[t!h!]
\includegraphics[width=0.8\textwidth]{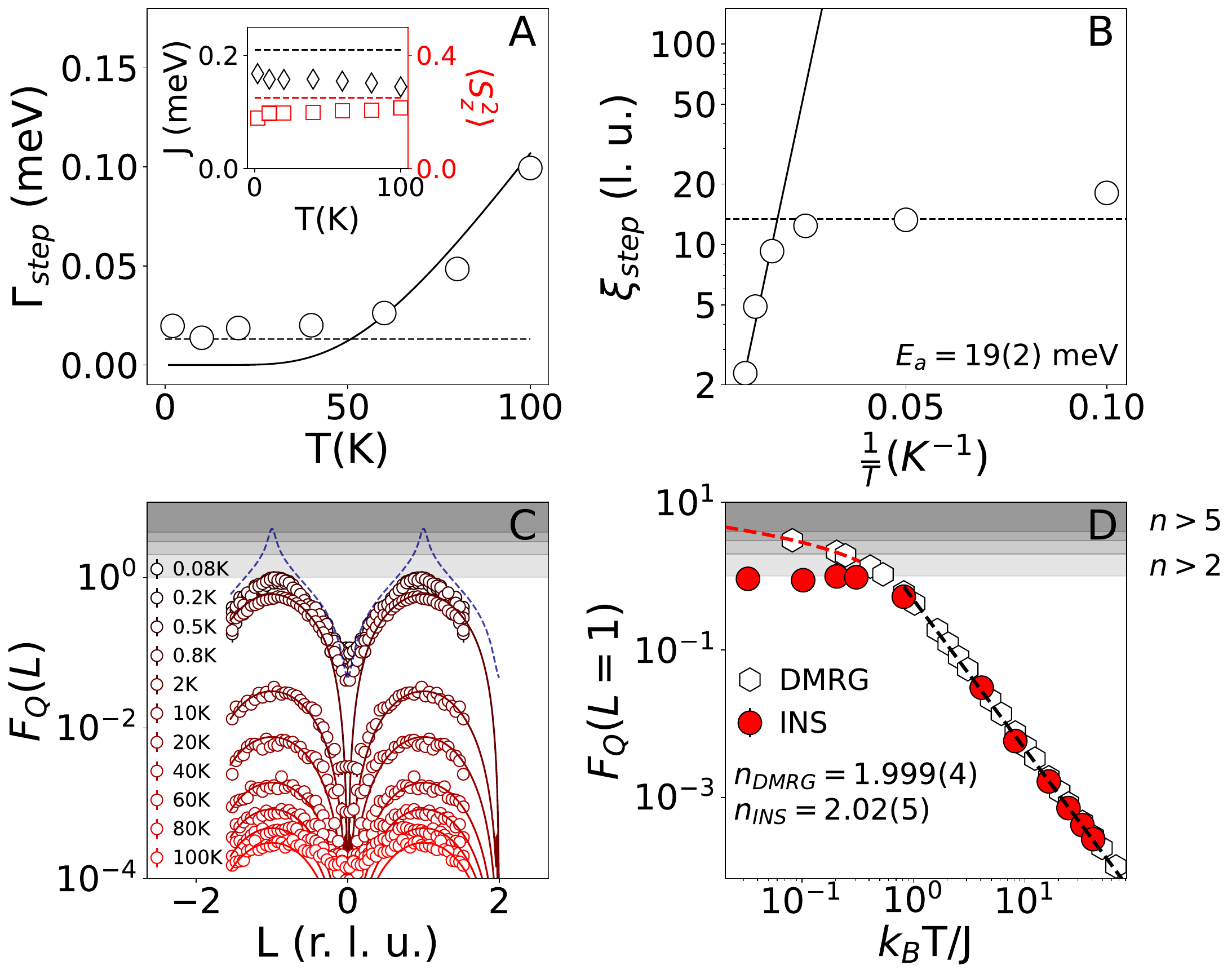}
\caption{
{\bf Temperature dependence of INS spectral parameters and quantum Fisher information ($F_Q$), using phenomenological step-function model.} (A) Life-time broadening parameter as a function of temperature obtained from step-function fits. Dashed line is instrumental resolution HWHM calculated at 0.5 meV. Solid curve is a fit to Arrhenius-type exponential function as described in the text. The inset shows fitted exchange interaction, $J$, and integrated intensity, $\langle S_z^2 \rangle$, at different temperatures; horizontal lines indicate nominal values, $J = 0.21$~meV \cite{Wu_2019,Nikitin_PRB2020} and $\langle S_z^2 \rangle = 1/4$. (B) Coherence length calculated using the spinon dispersion and extracted lifetime. Solid and dashed lines are asymptotic Arrhenius and resolution-limited behaviors as in (A). (C) Wave-vector dependence of the QFI, $F_Q(L)$, at various temperatures. Dashed curve is an approximation to asymptotic zero-temperature limit calculated from DMRG data at 200~mK as described in the text. (D) Temperature dependence of maximal QFI, $F_Q(L=1)$. Dashed black line is a power-law fit to the data in $T\geq 2$~K range capturing asymptotic high-temperature behavior, $F_Q \sim (J/T)^n$, with $n = 2$. Dashed red curve, shown in the region below $T_N=0.8$~K, is a fit of DMRG data below 4~K to a logarithmic dependence, $F_Q = \frac{3}{2} \ln({aJ}/{k_BT})$ with $J=0.21$~meV and fitting parameter $a=0.82$, illustrating the low-T asymptotic behavior; in YbAlO$_{3}$ it is arrested by static order below $T_N$, where part of the excitation spectrum condenses into elastic Bragg peaks that do not contribute to QFI.
}
\label{S_Fig7:t_pars_stepfnmodel}
\end{figure*}

\end{document}